\documentclass[12pt,reprint,tightenlines,groupedaddress,superscriptaddress,floatfix,aps]{revtex4}
\usepackage{graphicx}
\usepackage{rotating}
\usepackage{array}
\usepackage{multirow}
\usepackage{epstopdf}
\usepackage{amsmath}
\usepackage{amsfonts}
\usepackage{setspace}
\usepackage{braket}
\usepackage{color}

\usepackage{hyperref}

\DeclareMathOperator{\Tr}{Tr}

\begin{document}
\title{Broken-symmetry self-consistent GW approach: degree of spin contamination and evaluation of effective exchange couplings in solid antiferromagnets}

\author{Pavel Pokhilko}
\affiliation{Department  of  Chemistry,  University  of  Michigan,  Ann  Arbor,  Michigan  48109,  USA}
\author{Dominika Zgid}
\affiliation{Department  of  Chemistry,  University  of  Michigan,  Ann  Arbor,  Michigan  48109,  USA}
\affiliation{Department of Physics, University of Michigan, Ann Arbor, Michigan 48109, USA }

\renewcommand{\baselinestretch}{1.0}

\begin{abstract}
We adopt a broken-symmetry strategy for evaluating effective magnetic constants $J$ within 
the fully self-consistent GW method. 
To understand the degree of spin contamination present in broken-symmetry periodic solutions, 
we propose several size-extensive quantities 
that demonstrate that the unrestricted self-consistent GW preserves well 
the broken-symmetry character of the unrestricted Hartree--Fock solutions.  
The extracted $J$ are close to the ones obtained from multireference wave-function calculations. 
In this paper, we establish a robust computational procedure for finding magnetic coupling constants from self-consistent GW calculations and apply it to solid antiferromagnetic nickel and manganese oxides.
\end{abstract}
\maketitle

\section{Introduction}
Numerous strongly correlated electronic states, containing multiple important determinants 
in their wave-function expansions, 
are frequently necessary to describe magnetic phenomena\cite{White:magnetism:2007} and catalysis\cite{Boreskov:catalysis:2003}. 
While the wave-function methods\cite{OlsenText,Szabo_ostlund} provide a pathway for 
a systematic description of such states, 
the main limitation of these approaches remains in the substantial computational cost of building a determinantal expansion.
Consequently, wave-function methods are often too costly to perform solid-state calculations and to achieve a convergence to the thermodynamic limit (TDL).

As a cheaper alternative, density functional theory\cite{Kohn:64:DFT,Kohn65_DFT,Parr:Weitao:DFT:1994} (DFT) 
is the most commonly employed method for solids. 
For strongly correlated systems, within the unrestricted DFT, it is possible to produce several self-consistent solutions that display symmetry breaking.
These solutions can be used to reconstruct the structure of strong correlation and 
relate it to experimental observables\cite{noodleman:BS:81,Yamaguchi:BS:formulation:1986}. 
Despite its low computational cost, the main drawback of DFT lies in its lack of systematic improvability and 
in an ambiguous choice of density functional. 

As another alternative, Green's function methods\cite{Mahan00,Negele:Orland:book:2018,Martin:Interacting_electrons:2016} 
can be used to describe magnetic properties. 
The Green's function methods combine advantages of both DFT and the wave-function methods. 
The one-electron Green's function is a compact object in comparison with the wave-function amplitudes, 
reducing memory footprint necessary for computation. 
Any physical observable can be described using Green's functions, 
which led to numerous applications of Green's functions to multiple areas of condensed matter physics, such as  
transport phenomena, photoelectron spectroscopy\cite{Almbladh:photoemission:1985,Hedin:photoemission:1985,Fujikawa:photoelectron:chapter:2015}, and superconductivity\cite{Kadanoff:superconductivity:1961}.  
While in DFT the functional form of the density functional is unknown, 
the functional dependence of grand potential $\Omega[G]$ 
on the one-electron Green's function is known and can be systematically approximated\cite{Luttinger60}. 
Perturbative approximations of this functional give skeleton (bold) series, 
forming the theoretical basis of the self-consistent conserving Green's function methods\cite{Luttinger60,Baym61,Baym62}.  

Molecular non-relativistic calculations often use an $\braket{\hat{S}^2}$ observable as a 
diagnostic tool indicating the calculations' quality\cite{Szabo_ostlund}. 
The $\hat{S}^2$ operator commutes with the Hamiltonian; 
therefore, a common set of eigenfunctions exists for both electronic Hamiltonian and $\hat{S}^2$. 
However, approximate electronic states do not have to be eigenfunctions of $\hat{S}^2$ and
the expectation value $\braket{\hat{S}^2}$ evaluated for these approximate states 
does not have to be equal to any eigenvalue of the $\hat{S}^2$. 
This phenomenon is called a spin contamination\cite{note:spin_cont}. 
The deviation of the $\braket{\hat{S}^2}$ expectation value from the true eigenvalue of the $\hat{S}^2$ is 
a measure of spin contamination that is used for assessing how far the approximate description of 
the electronic state is from the spin-pure one. 
Evaluation of $\braket{S^2}$ is more useful than a simple diagnostic since  
various procedures for extraction of the effective magnetic couplings from the broken-symmetry DFT solutions rely 
on estimates of $\braket{\hat{S}^2}$, degrees of spin contamination, and orbital overlaps\cite{noodleman:BS:81,Yamaguchi:BS:formulation:1986,Malrieu:spin_pol:BS-DFT:2020,Malrieu:decont:BS-DFT:2020}. 

In solids, the phenomenon of spin contamination is much less analyzed. 
Solid-state calculations have a number of complications in comparison to molecular calculations. 
First, the number of transition metal centers can be very large, 
making a single $\braket{S^2}$ value less informative. 
Second, $\braket{S^2}$ is not an extensive property complicating  the analysis of spin contamination. 
Third, most DFT calculations lack a rigorous estimate of $\braket{S^2}$ 
since it is a two-electron property requiring a two-electron density matrix\cite{Cremer:DFT:S2:2001,Handy:DFT:S2:2007,Vedene:DFT:S2:1995}.  
Only very recently, addressing the first two concerns in the context of solids, a correction for spin contamination based on 
local spin--spin correlation function has been introduced in broken-symmetry DFT calculations\cite{Yamaguchi:APDFT:solid:2021}. 
However, even this correction depends on the definition of local spin. 

In solids, the evaluation of exchange coupling constants has a long history. 
Simple models and experimental data led to the formulation of rules and concepts, 
such as Goodenough--Kanamori rule\cite{Goodenough:book:1963,Kanamori:exchange_mechanisms:1959} and 
Ruderman--Kittel--Kasuya--Yosida interaction\cite{Kittel:RKKY:1954,Kasuya:RKKY:1956,Yosida:RKKY:1957}. 
When numerical computations became widely available, broken-symmetry Hartree--Fock and DFT calculations became 
the most common methods for evaluating effective magnetic constants\cite{Moreira:pccp_review:2006}. 
The wave-function strategies for evaluating effective exchange constants are usually 
performed by analyzing finite clusters\cite{Martin:NiO:exchange:2002,deGraaf:MRPT:exchange:solids:SMM:2001,Pokhilko:spinchain}.  
There are fewer strategies for evaluating effective exchange couplings within the Green's function calculations. 
Directly, in the solid regime, such an evaluation was performed in Ref.~\cite{Kotani:SF-GW:2008} by Kotani who employed the periodic quasiparticle self-consistent GW. 
We would like to note that the approach developed  by us, and presented here,  
is significantly different from the approach used in Ref.~\cite{Kotani:SF-GW:2008}  
 belonging to the spin-flip (SF) family of methods\cite{Casanova:SFReview} based on transversal spin susceptibility. The SF methods do not use broken-symmetry solutions.  
In our approach, we do not rely on a quasiparticle approximation employed in Ref.~\cite{Kotani:SF-GW:2008}  and we do not apply any additional approximations used there, such as independence of effective interaction $\tilde{U}$ from momentum, 
diagonal approximation for $\tilde{U}$, series expansions for effective interactions, 
and empirically parametrized muffin-tin potentials.

In our previous publications\cite{Pokhilko:tpdm:2021,Pokhilko:local_correlators:2021}, 
we evaluated $\braket{S^2}$ within self-consistent GF2 and GW for molecules, 
analyzed various self-consistent solutions, and formulated a broken-symmetry approach for evaluating
effective magnetic exchange couplings for both self-consistent GF2 and GW. 
In this work, we extend this development to solids. 
We propose two new extensive measures of spin contamination, which we denote as $SS_u^{{\bf k}=0}$ and $SS_u$, 
and compare them with the total $\braket{S^2}$ value and local spin--spin correlators. 
We analyze the impact of electronic cumulant (a connected two-particle contribution) 
and show that the approximate spin projection schemes based only on the disconnected contributions to $\braket{S^2}$ 
can overestimate spin contamination. 
We apply the formulated broken-symmetry self-consistent GW approach to solid nickel (NiO)  and manganese (MnO) oxides. 
The effective exchange couplings evaluated based on self-consistent GW for solid NiO and MnO agree well with the multireference wave-function estimates obtained from finite cluster calculations.
We also analyze the impact of the finite-size effects on spin contamination and the effective exchange constants. 

\section{Theory}
\subsection{Green's functions}
We work with imaginary-time one-particle Green's functions defined as an equilibrium expectation value
\begin{gather}
G_{pq} (\tau) = -\frac{1}{Z} \Tr \left[e^{-(\beta-\tau)(\hat{H}-\mu \hat{N})} p e^{-\tau(\hat{H}-\mu \hat{N})} q^\dagger  \right], \\
Z = \Tr \left[ e^{-\beta(\hat{H}-\mu \hat{N})} \right], 
\end{gather}
where $Z$ is a grand canonical partition function, 
$\hat{N}$ is a particle-number operator, 
$\mu$ is a chemical potential, 
$\tau$ is an imaginary time, 
$\beta$ is an inverse temperature, 
$p$ and $q^\dagger$ are annihilation and creation operators, respectively, expressed in a spin-orbital basis. 
The antiperiodicity of the Green's function results in an odd (fermionic) discrete set of Matsubara frequencies $\omega_n$, 
while symmetric in time periodic quantities are expressed through even (bosonic) Matsubara frequencies $\Omega_n$
\begin{gather}
\omega_n = \frac{(2n+1) \pi}{\beta}, \\
\Omega_n = \frac{2n \pi}{\beta}. 
\end{gather}
The correlated Green's function $G(i\omega_n)$ is expressed in terms of the  zeroth-order Green's function $G_0(i\omega_n)$ and self-energy via the Dyson equation written as
\begin{gather}
G^{-1}(i\omega_n) = G^{-1}_0(i\omega_n) - \Sigma[G](i\omega_n),
\protect\label{eq:Dyson}
\end{gather}
where $\Sigma[G](i\omega_n)$ is the self-energy, 
depending on the full Green's function as a functional. 
Particular approximations define the structure of this functional dependence. 
Note that the self-energy can be further separated into
\begin{gather}
\Sigma[G](i\omega_n)=\tilde{\Sigma}[G](i\omega_n)+\Sigma[G],
\end{gather}
where the first part $\tilde{\Sigma}[G](i\omega_n)$ is the dynamic part of the self-energy, while the second part defines the static (frequency independent) part of the self-energy $\Sigma[G]$.

In this paper, we use the unrestricted Hartee--Fock approximation (UHF) 
and the unrestricted self-consistent finite-temperature GW approximation~\cite{Hedin65,G0W0_Pickett84,G0W0_Hybertsen86,GW_Aryasetiawan98,Stan06,Koval14,scGW_Andrey09,Kutepov17,Iskakov20,Yeh:GPU:GW:2022}.  
The Hartree--Fock approximation employs only the static part of self-energy.
Correlated self-consistent Green's function methods use a skeleton diagram expansion to approximate both the static and dynamic parts of the self-energy.
In particular, the GW approximation treats electronic correlation through bubble diagrams (terms) that can be expressed through an effective screened frequency-dependent interaction $W(\Omega_n)$. 

Using the thermodynamic Hellmann--Feynman theorem, we demonstrated in our recent work~\cite{Pokhilko:tpdm:2021} that the two-particle density matrices 
within the self-consistent Green's function methods decompose into 
a disconnected contribution $\Gamma^\text{disc}$ and 
a connected contribution $\Gamma^\text{conn}$ (also called an electronic cumulant)
\begin{gather}
\Gamma_{\braket{p q |r s}} = 
\Gamma^\text{disc}_{\braket{p q |r s}} + \Gamma^\text{conn}_{\braket{p q |r s}}, \\
\Gamma^\text{disc}_{\braket{p q |r s}} = \gamma_{pr} \gamma_{qs} - \gamma_{ps} \gamma_{qr},  
\protect\label{eq:Gamma}
\end{gather}
where $\gamma_{pq} = -G_{qp}(\beta^-)$ is the correlated one-particle density matrix evaluated from the correlated Green's function.
The 4-index conventions used here correspond to the electron repulsion integrals (ERI): 
$\braket{\cdots}$ denotes ERI in the physical notation and $(\cdots)$ denotes ERI in the chemical notation, 
defined as 
\begin{gather}
\braket{pq | rs} = (pr | qs) = 
\int \phi_p^*(\mathbf{r}_1;\sigma_1)\phi_q^*(\mathbf{r}_2;\sigma_2) \frac{1}{|\mathbf{r}_1-\mathbf{r}_2|} 
\phi_r(\mathbf{r}_1;\sigma_1)\phi_s(\mathbf{r}_2;\sigma_2) d\mathbf{r}_1 d\mathbf{r}_2 d\sigma_1  d\sigma_2.
\end{gather}

The Hartree--Fock method equates the electronic cumulant to zero, 
explaining that there is no Coulomb correlation between electrons within the Hartree--Fock approximation. 
The GW cumulant is expressed as a product between the screened interaction and polarization bubbles $\Pi(\Omega_m)$
\begin{gather}\label{eq:gw_cummulant}
\Gamma_{(p_0 q_0 | r_0 s_0)}^\text{GW \ conn} = 
-\frac{1}{\beta}\sum_{\Omega_m} 
\sum_{pqrs} \Pi_{r_0 s_0 pq}(\Omega_m) W_{(pq|rs)}(\Omega_m) \Pi_{rs p_0 q_0}(\Omega_m). 
\end{gather}
The two-particle density matrices derived in this manner fully reproduce the two-body energy, 
making this approach convenient for analysis of energies and resulting electronic structure. 
The dynamical properties~\cite{Iskakov:phase_tr:2022} can be evaluated within this approach as well.

\subsection{One- and two-particle properties}
One- and two-particle properties are evaluated from the corresponding one- and two-particle density matrices. The one- and two-particle operators are respectively  defined as 
\begin{gather}
\hat{B} = \sum_{pq} b_{pq} p^\dagger q \\
\hat{C} = \sum_{pqrs} c_{\braket{pq|rs}} p^\dagger q^\dagger s r. 
\end{gather}
The respective expectation values of the one- and two-particle operators then take the form
\begin{gather}
\braket{\hat{B}} = \sum_{pq} b_{pq} \braket{p^\dagger q} = \sum_{pq} b_{pq} \gamma_{pq} 
\protect\label{eq:property_trace1}
\end{gather}
\begin{gather}
\braket{\hat{C}} = \sum_{pqrs} c_{\braket{pq|rs}} \braket{p^\dagger q^\dagger s r} 
= \sum_{pqrs} c_{\braket{pq|rs}} \Gamma_{\braket{pq|rs}}.
\protect\label{eq:property_trace2}
\end{gather}

In this paper, we focus on analyzing spin properties of solutions obtained within the Green's function formalism, and here specifically GW. 
Both one- and two-particle properties evaluated for each of these solutions provide deep insight into their electronic structure. 

In practice, it can be easier to rationalize results of calculations 
based on certain one- and two-particle properties obtained as a trace with the corresponding integrals, 
as shown in Eqs.\ref{eq:property_trace1}-\ref{eq:property_trace2}.  
In particular, $\braket{S^2}$, which is an expectation value of a two-particle operator, is commonly used for the assessment of quality of quantum chemical calculations of open-shell species---systems with unpaired electrons\cite{Amos:spin_cont:1991,Schlegel:S2:94,Schlegel:spin_cont:1998,Stanton:CCSD_S2:1994,Krylov:S2:CC:2000}. 
The summation in the Eq.\ref{eq:property_trace2} can be performed only over a subset of orbitals, 
providing, for example, local spin--spin and charge correlators, analysis of magnetic properties\cite{Davidson:local_spin:2001,Davidson:local_spin:2002,Davidson:MolMagnets:2002,Hess:local_spin:2005,Pokhilko:local_correlators:2021}, chemical bond orders\cite{Ruedenberg:chem_bond:1962,Jorge:bond_index:1985,Torre:popul:cumulants:2002,Bochicchio:bond_order:2003,Goddard:corr_chem_bond:1998,Luzanov:bond_indices:2005,Mayer:bond_order:2007}, and charge transfer\cite{Luzanov:SpinCorr:15,Pokhilko:local_correlators:2021}. 

Previously, in Refs.~\onlinecite{Pokhilko:tpdm:2021,Pokhilko:local_correlators:2021}, we have used global $\braket{S^2}$ and $\braket{N^2}$ diagnostics as well as 
local spin and charge correlators 
for characterizing and benchmarking the Green's function solutions that appear in the broken-symmetry self-consistent GW approach. These solutions were subsequently used to evaluate effective magnetic couplings\cite{Pokhilko:local_correlators:2021}. 
Here, we generalize such an approach to periodic solids and 
apply it to understand the origin of the effective exchange in periodic transition metal oxides.

We work with periodic Bloch orbitals defined as 
\begin{gather}
\phi_{\mathbf{k},i} = \sum_{\mathbf{R}} \phi_i^{\mathbf{R}} (\mathbf{r}) e^{i\mathbf{k}\cdot\mathbf{R}},
\end{gather}
where $\phi_i^{\mathbf{R}} (\mathbf{r})$ is an atom-centered combination of primitive Gaussian functions, 
$\mathbf{R}$ is a Bravais lattice cell, $\mathbf{k}$ is the momentum vector index. 
Within this approach, a compact Gaussian basis set is used for expressing both occupied and unoccupied orbitals. In such a basis, the number of unoccupied orbitals remains relatively low resulting in a relatively small total number of orbitals (occupied and unoccupied) in each unit cell.  This is 
important for reducing the overall computational cost of correlated calculations. 

A generalization of the computational expressions to periodic systems is rather straightforward 
and is obtained by replacing the molecular spin-orbital index $i$ with a multi-index $\mathbf{i} = (i,\mathbf{k})$ 
and applying the momentum conservation 
(which is analogous to use of symmetry adaptation within Abelian point groups in molecular calculations). 
Due to the momentum conservation, only one momentum index is necessary for storing the one-particle Green's function 
and three momenta indices are required for storing 4-index quantities, such as ERI and $W$. 
The details of our periodic GW formulation are outlined in the Ref.\cite{Iskakov20,Yeh:GPU:GW:2022} and will not be repeated here. 

In the evaluation of correlators, a special consideration should be given to the contractions between density matrices and integrals.
In a periodic implementation, the one- and two-body energies are evaluated as
\begin{gather}
E_{1b} = \frac{1}{V_{BZ}}\sum_{pq} h_{pq}^\mathbf{k} \gamma_{pq}^\mathbf{k}, \\
E_{2b} = \frac{1}{2V_{BZ}}\sum_{\omega_n} \sum_{pq} G_{qp}^\mathbf{k}(\omega_n) \Sigma_{pq}^\mathbf{k}(\omega_n), 
\end{gather}
where $V_{BZ}$ is the volume of the Brillouin zone. 
Here, the second expression is the Galitskii--Migdal formula for the two-body energy following from the Heisenberg equation of motion for creation and annihilation operators. 
Note that both expressions are equipped with $1/V_{BZ}$ prefactors, 
meaning that the energy is evaluated per primitive cell.
The numerical recipes for evaluating $V_{BZ}$ can differ due to inclusion of symmetries; 
in the simplest case when all types of symmetry except translations are neglected, 
$V_{BZ}$ is equal to the total number of k-points.

Since self-consistent GW is a conserving approximation, the thermodynamic consistency of  
energy is satisfied and the two-particle density matrix reproduces the two-body energy
\begin{gather}
E_{2b} = \frac{1}{2} \frac{1}{V^2_{BZ}}\sum_\mathbf{pqrs} \braket{\mathbf{pq}|\mathbf{rs}} \Gamma_{\braket{\mathbf{pq}|\mathbf{rs}}},
\end{gather}
where the explicitly stored momenta indices are not shown for brevity. 
Note that in the last expression the prefactor is $1/V^2_{BZ}$. 

The evaluation of properties {\em per cell } can be done only with extensive properties, such as energies. 
Total $\braket{S^2}$ is not an extensive property; 
therefore, the multiplications by $1/V_{BZ}$ or $1/V^2_{BZ}$ are not present. 
The  $\braket{S^2}$  expressions are given in SI of Ref.~\cite{Pokhilko:tpdm:2021} and
can be fully re-used with the multi-index generalization, explicitely shown in Section~\ref{sec:S2_expressions}. 
The resulting quantity shall be interpreted as $\braket{S^2}$ of the corresponding supersystem (for all the cells, not just the unit cell) in the real space. Knowing such a defined $\braket{S^2}$
offers a direct comparison with molecular calculations.  

In periodic calculations, it is necessary to investigate the convergence of evaluated quantities to the thermodynamic limit (TDL).  At TDL, a central unit cell is surrounded by sufficiently many cells such that their presence cease to influence the local extensive properties evaluated within this cell.
In solids, due to periodicity, the convergence to TDL is considered either
with respect to the system size in the real space or 
with respect to the number of k-points in the reciprocal space. 
Since both the real space and k-space are connected via the Fourier transform, it is sufficient to consider the convergence only within one of them. 
Due to the computational setup of our calculations, 
we investigate the convergence with respect to the number of k-points.

Since total $\braket{S^2}$ is not an extensive property, its change with the number of k-points is not proportional to their number.
Consequently, other extensive quantities that reflect the degree of spin contamination are more convenient for numerical analysis within the unit cell. 
We propose two of such quantities: $SS_u^{\mathbf{k}=0}$ and $SS_u$.

\subsubsection{Expressions for $\braket{S^2}$ and related quantities} 
\protect\label{sec:S2_expressions}
We use AOs, which are non-orthogonal atomic orbitals. 
We use the covariant notation, in which the second-quantized operators satisfy 
the following anticommutation relations
\begin{gather}
\{ \mathbf{p}^\dagger, \mathbf{q}^\dagger \} = 
\{ \mathbf{p}, \mathbf{q} \} = 0 \\
\{ \mathbf{p}^\dagger, \mathbf{q} \} = S^{-1}_{\mathbf{pq}} = (S^{\mathbf{k}})^{-1}_{pq} \delta_{\mathbf{k}_p\mathbf{k}_q},
\end{gather}
where  $\mathbf{p}^\dagger$ and $\mathbf{q}$ are the creation and annihilation operators for the Bloch orbitals, 
$S_\mathbf{pq}$ is the overlap matrix between the Bloch orbitals, $\mathbf{k}$ is a momentum vector. 

Spin ladder operators and $S^2$ are
\begin{gather}
S_+ = S_x + iS_y, \\
S_- = S_x - iS_y, \\
S^2 = S_x^2 + S_y^2 + S_z^2 = 
S_-S_+ +  S_z + S_z^2. 
\end{gather}
In the derivation, we assume ensemble representability, 
allowing us to write thermal averages with brakets $\braket{...}$. 
The expressions for the spin operators in Bloch spin-orbitals are
\begin{gather}
S_\mu = \sum_\mathbf{pq} \braket{\mathbf{p}|S_\mu|\mathbf{q}} \mathbf{p}^\dagger \mathbf{q}, \\
S_\mu S_\nu = \sum_\mathbf{pqrs} \braket{\mathbf{p}|S_\mu|\mathbf{q}} \braket{\mathbf{r}|S_\nu|\mathbf{s}} \mathbf{p}^\dagger \mathbf{q} \mathbf{r}^\dagger \mathbf{s} = \nonumber \\
-\sum_\mathbf{pqrs} \braket{\mathbf{p}|S_\mu|\mathbf{q}} \braket{\mathbf{r}|S_\nu|\mathbf{s}} \mathbf{p}^\dagger \mathbf{r}^\dagger \mathbf{q  s} + \nonumber \\
\sum_\mathbf{pqrs} \braket{\mathbf{p}|S_\mu|\mathbf{q}}S_\mathbf{qr}^{-1} \braket{\mathbf{r}|S_\nu|\mathbf{s}} \mathbf{p}^\dagger  \mathbf{s}.
\end{gather}
In atomic orbitals, the expressions are
\begin{gather}
\braket{S_-S_+} = 
-\sum_\mathbf{pqrs} 
S_\mathbf{pq} S_\mathbf{rs}\Gamma_{\braket{\mathbf{pr}|\mathbf{sq}}}^{\beta\alpha \beta \alpha}  + 
\sum_\mathbf{pqrs} S_\mathbf{pq} S^{-1}_\mathbf{qr} S_\mathbf{rs} \gamma_\mathbf{ps}^{\beta\beta}
\protect\label{eq:SmSp} \\
\braket{S_+S_-} = 
-\sum_\mathbf{pqrs} 
S_\mathbf{pq} S_\mathbf{rs} \Gamma_{\braket{\mathbf{pr}|\mathbf{sq}}}^{\alpha\beta \alpha \beta}  + 
\sum_\mathbf{pqrs} S_\mathbf{pq} S^{-1}_\mathbf{qr} S_\mathbf{rs}  \gamma_\mathbf{ps}^{\alpha\alpha} 
\protect\label{eq:SpSm} \\
\braket{S_z^2} = 
-\frac{1}{4} \sum_\mathbf{pqrs} S_\mathbf{pq} S_\mathbf{rs} (
\Gamma_{\braket{\mathbf{pr}|\mathbf{sq}}}^{\alpha\alpha\alpha\alpha}+
\Gamma_{\braket{\mathbf{pr}|\mathbf{qs}}}^{\alpha\beta\alpha\beta}+
\Gamma_{\braket{\mathbf{pr}|\mathbf{qs}}}^{\beta\alpha\beta\alpha}+
\Gamma_{\braket{\mathbf{pr}|\mathbf{sq}}}^{\beta\beta\beta\beta}) \nonumber \\
+ \frac{1}{4} \sum_\mathbf{pqrs} S_\mathbf{pq} S^{-1}_\mathbf{qr} S_\mathbf{rs} 
(\gamma_\mathbf{ps}^{\alpha\alpha} + \gamma_\mathbf{ps}^{\beta\beta}) 
\protect\label{eq:SzSz} \\
\braket{S_z} = 
\frac{1}{2}\sum_\mathbf{pq} S_\mathbf{pq} (\gamma_\mathbf{pq}^{\alpha\alpha}-\gamma_\mathbf{pq}^{\beta\beta})
\protect\label{eq:Sz}
\end{gather}
The expressions for $SS_u^{\mathbf{k}=0}$, $SS_u$, and $SS_{AB}$ are generated by 
performing the summations above over a restricted set of momenta ($SS_u^{\mathbf{k}=0}$, $SS_u$, $SS_{AB}$) 
and orbitals ($SS_{AB}$) and a final division over $V_{BZ}$ (only for $SS_u$ and $SS_{AB}$). 
While for  total $\braket{\hat{S}^2}$ the expressions can be further simplified since $S^{-1}$ cancels out, 
this does not happen when the summation runs over restricted sets of spin-orbitals.

\subsubsection{Evaluation of $SS_u^{\mathbf{k}=0}$ }\label{sec:SSuk}
For $SS_u^{\mathbf{k}=0}$, we take only contributions from the $\Gamma$ point ($\mathbf{k} = 0$) 
\begin{gather}
SS_u^{\mathbf{k}=0} = \braket{\vec{S}_u^{\mathbf{k}=0} \vec{S}_u^{\mathbf{k}=0}},
\end{gather}
where $S_u^{\mathbf{k}=0}$ denotes the overlap integrals $S_{pq}^{\mathbf{k}=0}$ for all orbitals $p,q$ 
in the unit cell with $\mathbf{k}=0$, 
which are summed over with density matrices to yield $SS_u^{\mathbf{k}=0}$. 
The vector symbol $\vec{S}$ is a shortcut for a full expression involving $S_+$, $S_-$, and $S_z$. 
This approach is very simple and fully re-uses the molecular code for $\braket{S^2}$, 
since this approach constrains the summations in Eqs.~\ref{eq:SmSp}--\ref{eq:Sz} 
over $\mathbf{p},\mathbf{q}, \mathbf{r}, \mathbf{s}$ to run only over the Bloch orbitals with $\mathbf{k}=0$.  

Unfortunately, evaluating the electronic cumulant from Eq.~\ref{eq:gw_cummulant} is very expensive and currently possible only for very few k-points. 
Therefore, to test the impact of the cumulant, we include only the cumulant with 4 zero momentum indices in the $SS_u^{\mathbf{k}=0}$ diagnostic. 
Neglecting the cumulant has another reasoning as well. 
We showed that perturbative Green's function methods 
produce a non-zero $\braket{S^2}$
even for closed-shell systems\cite{Pokhilko:tpdm:2021}. 
The origin behind such a spin contamination lies in the partial inconsistency in the perturbative orders of the 
density matrices entering the computational expressions of $\braket{S^2}$. 
In GW, most of this spin contamination comes from the connected contribution. Consequently, 
neglecting the cumulant gives better $\braket{S^2}$ estimates for single solutions.  
However, in the Ref.~\cite{Pokhilko:local_correlators:2021}, we showed that 
when the differences in $\braket{S^2}$ between different solutions are considered, 
this artificial spin contamination mostly cancels out. 
The differences in $\braket{S^2}$ are important because they are frequently used within the 
Yamaguchi's correction for spin contamination for effective exchange couplings. 
In this work, we evaluate the differences in $SS_u^{\mathbf{k}=0}$ and $SS_u$ and 
show that the analogous statement holds true in solids as well.

\subsubsection{Evaluation of $SS_u$ }\label{sec:SSu}
For $SS_u$, we sum over the entire momentum space as follows
\begin{gather}
SS_u = \frac{1}{V_{BZ}}\sum_\mathbf{k} \braket{\vec{S}_u^{\mathbf{k}} \vec{S}_u^{\mathbf{k}}}.
\end{gather}
This approach corresponds to evaluating a spin--spin correlation function 
for which we also neglect the electronic cumulant. 
Neglecting the cumulant is performed in nearly all practical estimations of $\braket{S^2}$ within DFT, 
making the diagnostics described here also applicable to DFT calculations. 
$SS_u$ constrains the summations in the Eqs.~\ref{eq:SmSp}--\ref{eq:Sz} 
over $\mathbf{p},\mathbf{q}, \mathbf{r}, \mathbf{s}$ such that $\mathbf{k_p}=\mathbf{k_q}=\mathbf{k_r}=\mathbf{k_s}$.

\subsubsection{Evaluation of local spin correlators }
To gain additional insight into the Green's functions obtained, 
we evaluate local spin correlators considering only subsets of orbitals $A$ and $B$ 
\begin{gather}
SS_{AB}^{\mathbf{k}=0} = \braket{\vec{S}_A^{\mathbf{k}=0} \vec{S}_B^{\mathbf{k}=0}},  \\
SS_{AB} = \frac{1}{V_{BZ}}\sum_\mathbf{k} \braket{\vec{S}_A^\mathbf{k} \vec{S}_B^\mathbf{k}}. 
\end{gather}
For both correlators, the summations in Eqs.~\ref{eq:SmSp}--\ref{eq:Sz} are constrained over the momentum in the same manner as for $SS_u^{\mathbf{k}=0}$ and $SS_u$. 
The definition of the local correlators depends on the chosen sets $A$ and $B$. 
As in the previous paper\cite{Pokhilko:local_correlators:2021}, here
we use Mulliken partitioning, for which the orbital subsets are all atomic orbitals on a given center. 
Since we use covariant density matrices, 
evaluation of properties from  Eqs.~\ref{eq:property_trace1}--\ref{eq:property_trace2} 
preserves the trace in a non-orthogonal orbital basis. 
Such local correlators provide information about electronic correlation 
that is otherwise not easily explained based on the  one-electron observables. 
In solids, the local spin--spin correlators have been used previously to estimate spin contamination in DFT calculations\cite{Yamaguchi:APDFT:solid:2021}. 

In this work, we compare $SS_u^{\mathbf{k}=0}$, $SS_u$, and local spin--spin correlators, 
as well as analyze their ability to measure spin contamination.

\subsection{Effective magnetic Hamiltonians}
If a low-energy manifold of electronic states is separated from all other electronic states, 
an effective Hamiltonian describing only the targeted state manifold can be easily built.
In principle, such a description can be exact. 
A generalism Bloch formalism\cite{Bloch:1958,Cloizeax:1960,Okubo:1954} provides a 
theoretical foundation for construction of effective Hamiltonians, 
which became a platform for modern multireference methods. 
Effective Hamiltonians often have a simple structure making them useful for the analysis and interpretation of electronic structure. 
For example, effective magnetic Hamiltonians, such as the Heisenberg Hamiltonian, 
can be rigorously derived using Bloch formalism\cite{Calzado:02,Guihery:2009,Malrieu:2010,Marlieu:MagnetRev:2014,Mayhall:2014:HDVV,Mayhall:1SF:2015,Pokhilko:EffH:2020,Pokhilko:spinchain}. 
Moreover, effective Hamiltonians provide a ground for robust extrapolations, 
capturing electronic states 
that are not explicitely involved in the calculation\cite{Mayhall:1SF:2015,Pokhilko:spinchain}. 

\begin{figure}[!h]
  \includegraphics[width=8cm]{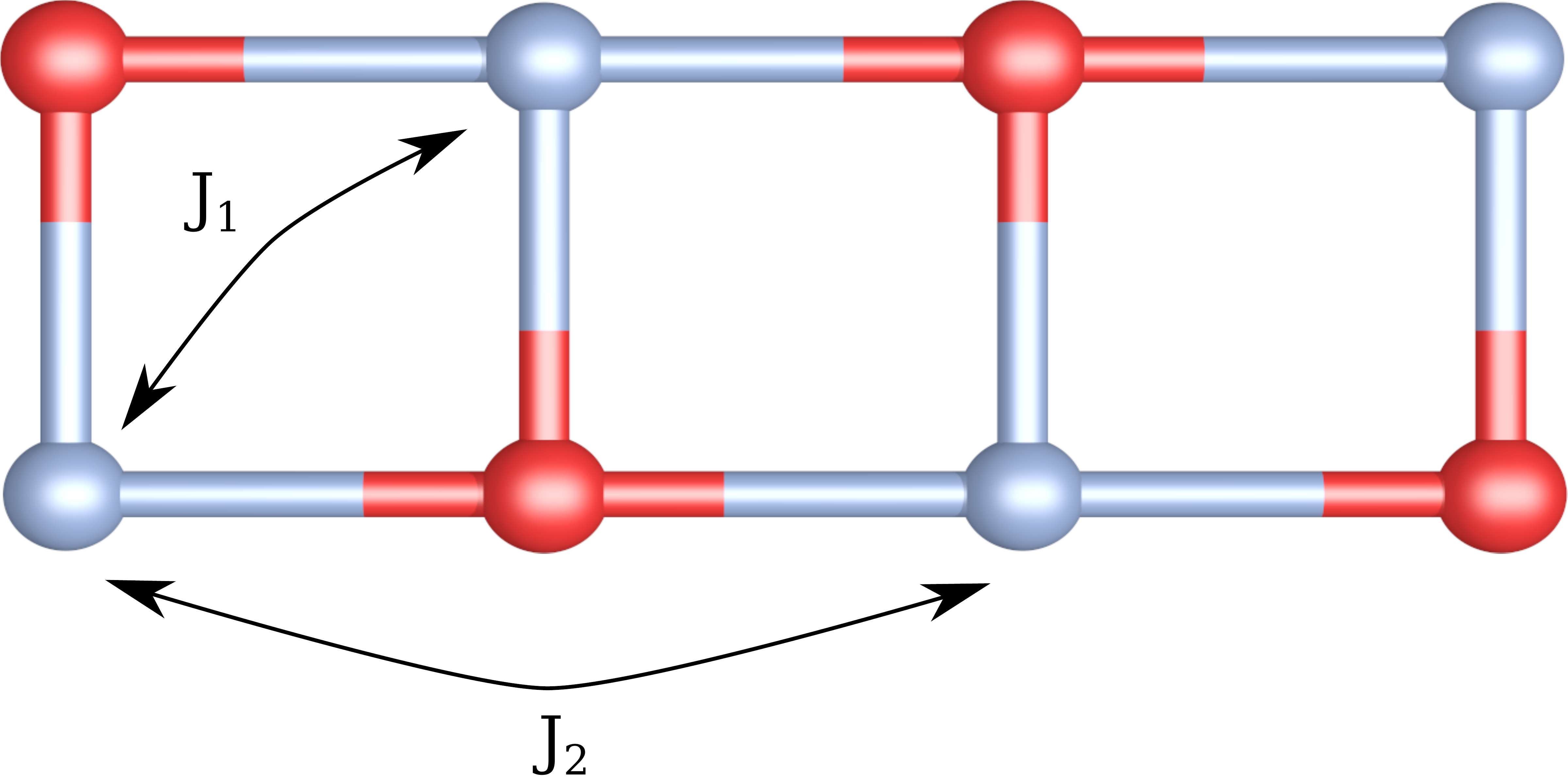}
\centering
\caption{Nearest-neighbor (NN) couplings $J_1$ and next-nearest-neighbor (NNN) couplings $J_2$. 
         Metal centers are shown in grey; oxygen atoms are shown in red. 
         \protect\label{fig:J_def}
}
\end{figure}
\begin{figure}[!h]
  \includegraphics[width=12cm]{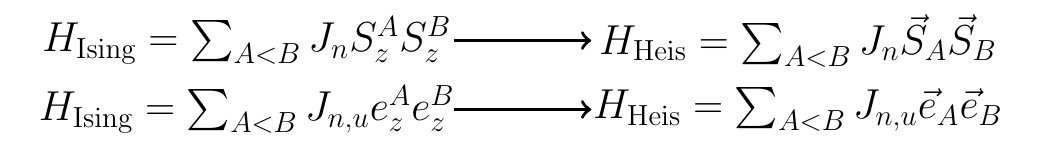}
\centering
\caption{Construction of the Heisenberg Hamiltonians from the Ising Hamiltonians. 
         \protect\label{fig:Ham_defs}
}
\end{figure}

Unfortunately, applications of the Bloch formalism require access to configurations and the wave-function amplitudes, 
which are not easily accessible from the Green's function calculations. 
Instead, in the broken-symmetry DFT approach, 
pioneered by Noodleman\cite{noodleman:BS:81} and Yamaguchi\cite{Yamaguchi:BS:formulation:1986}, 
it is possible to obtain high-spin and broken-spin (broken-symmetry) unrestricted solutions, 
interpret their energies as solutions of the Ising Hamiltonian, 
extract the effective exchange couplings, 
and construct the effective Heisenberg Hamiltonian (Figure~\ref{fig:Ham_defs}). 
Due to its simplicity,  
this approach is most commonly used for evaluating the effective exchange couplings in molecular\cite{Noodleman:BS:1986,Ruiz:BSDFT:99,Nishino:BSDFT:97,Soda:BSDFT:00,sinnecker:BS:Mn:04} and periodic\cite{Moreira:pccp_review:2006} magnets. 

The broken-symmetry DFT has been applied to MnO and NiO\cite{Martin:NiO:exchange:2002,Feng:MnO:CoO:b3lyp:2004,Kresse:MnO:PBE:2005,Majumdar:NiO:MnO:DFT:J:2011}. 
Both oxides have a rock-salt cubic crystal structure that becomes slightly distorted at low temperatures. 
It is possible to capture broken-symmetry solutions (BS) of two types\cite{note:BS-AFM}, labeled here as 
\textbf{type 1} and \textbf{type 2}. 
Within the Ising model, these solutions can be interpreted as antiferromagnetic phases of A-type, 
named as AF$_1$ and AF$_2$ in the Refs. \cite{Martin:NiO:exchange:2002,Majumdar:NiO:MnO:DFT:J:2011}. 
When applied to MnO and NiO, only two types of couplings are non-negligible and determine the magnetic ordering (Figure~\ref{fig:J_def}): 
$J_1$ is the coupling between nearest-neighbor metal centers (NN), connected via oxygen atoms (metal-oxygen-metal angle is 90$^\circ$);
$J_2$ is the coupling between the next-nearest-neighbor metal centers (NNN), connected via oxygen atoms (metal-oxygen-metal angle is 180$^\circ$).  
Consistently with previous papers\cite{Martin:NiO:exchange:2002,Majumdar:NiO:MnO:DFT:J:2011}, 
the magnetic Heiseberg Hamiltonian is
\begin{gather}
H = -J_{1,u} \sum_{\braket{i,j}} \vec{e}_i \vec{e}_j - J_{2,u} \sum_{\braket{\braket{i,j}}}  \vec{e}_i \vec{e}_j,
\protect\label{eq:Ham_defs}
\end{gather}
where the summation runs over unique pairs (there is no repetition in each of the sums) $\braket{i,j}$ denoting the NN and $\braket{\braket{i,j}}$ denoting the NNN. 
The unit spin form provides a convenient way of comparing J couplings for compounds with different local effective spins. 
If it is necessary not just to compare different compounds, but also work with their magnetic Hamiltonians 
(e.g., to obtain experimental observables), a conversion to the effective local spin $S$ is necessary. 
For NiO, the effective local spin is $1$---it coincides with the unit spin and $J_{n,u} = J_{n}$. 
For MnO, the effective local spin is $5/2$; 
therefore, the conversion follows from the different manner of writing the Heisenberg Hamitonian, 
$\braket{J_n \vec{S}_i \vec{S}_j} = \braket{J_{n,u} \vec{e}_i \vec{e}_j}$, and $J_{n} = \frac{4}{25} J_{n,u}$. 

Unfortunately, DFT is not systematically improvable
and the effective exchange couplings obtained depend on a chosen functional. 
Moreover, spin contamination of DFT solutions can limit applicability of the approach. 
To ease this limitation, a number of corrections based on spin projections, 
$\braket{S^2}$, and local spin--spin correlation functions 
have been introduced for molecules\cite{Yamaguchi:EHF:1988} and recently for solids\cite{Yamaguchi:APDFT:solid:2021}.  
In Refs.~\cite{Yamaguchi:EHF:1988,Yamaguchi:APUMP:1989,Yamaguchi:APCCSD:2012,Stanton:BS-CC:2020} it has been shown that weakly correlated wave-function methods often preserve the broken-symmetry nature of solutions, thus
providing reasonably accurate estimates of the 
effective exchange couplings.  
Symmetry breaking also happens in quasiparticle GW\cite{Loos:GW:discont:2018}. 
Recently, we demonstrated that self-consistent Green's function methods also 
preserve the broken-symmetry structure of the solutions\cite{Pokhilko:local_correlators:2021}, 
following the generalized Fukutome's classification of 
the Green's function solutions\cite{Mochena:Fukutome:broken_symmetry:GF}. 
Early benchmark calculations for single-molecule magnets\cite{Pokhilko:local_correlators:2021} indicate 
that accuracy of 
the formulated broken-symmetry self-consistent GW  approach is comparable to multireference wave-function methods 
and spin-flip equation-of-motion coupled-cluster method with single and double excitations (EOM-SF-CCSD). 

\section{Results and discussion}
\protect\label{sec:results}

\subsection{Computational details}
\begin{figure}[!h]
  \includegraphics[width=8cm]{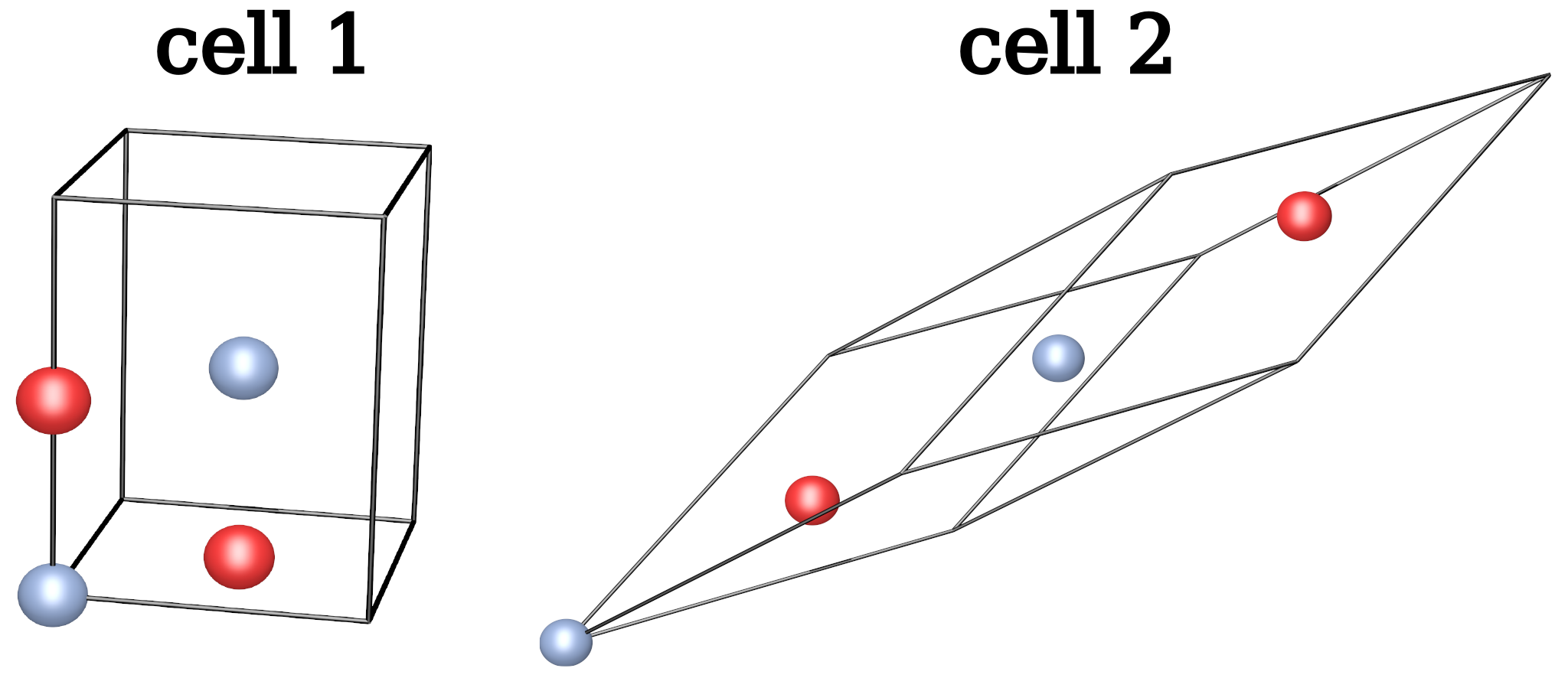}
\centering
\caption{Two unit cells used in calculations in this paper to capture solutions of different types. 
         Both unit cells produce the same primitive cell of rock salt type.
         \protect\label{fig:lattice}
}
\end{figure}

We used the same setup for periodic calculations as in  Ref.~\cite{Iskakov20}: 
\emph{gth-dzvp-molopt-sr} basis set\cite{GTHBasis}, \emph{gth-pbe} pseudopotential\cite{GTHPseudo}, and
\emph{def2-svp-ri} auxiliary basis\cite{RI_auxbasis} for
the resolution-of-identity decomposition. 
Since magnetic electronic states share similar occupancies of magnetic orbitals, 
the errors due to a finite basis set mostly cancel out when the coupling evaluation is based on the computation of differences.
It has been shown that double-zeta basis sets with polarization functions are usually sufficient for effective magnetic couplings\cite{Pavel:OSFNO:2019,Pokhilko:spinchain}.  
We used the NiO and MnO structures with the lattice constants a = 4.1705\AA~\cite{Morosin:NiO:exchange_striction:1971}
and 4.4450\AA~\cite{MnO_a}, respectively. 
To get ferromagnetic and antiferromagnetic solutions of different types, 
we used the unit cells doubled along different directions (Figure~\ref{fig:lattice}).
We used the Monkhorst--Pack k-point grid for the Brillouin-zone sampling\cite{Monkhorst:Pack:k-grid:1976}. 
To account for finite-size effects, 
the exchange term in GW is corrected by the Ewald approach\cite{EwaldProbeCharge,CoulombSingular}. 
As a frequency grid, we used an intermediate representation\cite{Yoshimi:IR:2017} grid
with $\Lambda = 10^5$ and 136 functions. 
The one- and two-electron integrals were computed with the PySCF code\cite{PYSCF}. 
The correlated calculations were performed within the resolution-of-identity approximation with 
the local in-house Green's function code \cite{Rusakov16,Iskakov20,Pokhilko:tpdm:2021,Pokhilko:local_correlators:2021}. 
To converge the GW calculations, we used the frequency-dependent CDIIS algorithm\cite{Pokhilko:algs:2022}.

For each of the cells (Figure~\ref{fig:lattice}), 
we converged zero-temperature UHF solutions with maximum spin projection (ferromagnetic solution, HS) 
and the broken-symmetry solutions (antiferromagnetic, BS) with zero spin projection.  
Using these zero-temperature solutions as guesses, 
we converged the finite-temperature UHF and unrestricted finite-temperature self-consistent GW calculations at $\beta = 500$~a.u.$^{-1}$. 
As we observed previously in molecular calculations, 
the properties of these solutions do not significantly depend on temperature and 
preserve the zero-temperature limit\cite{Pokhilko:local_correlators:2021}. 
We label the broken-symmetry solutions obtained in cells \textbf{1} and \textbf{2} as \textbf{BS1} and \textbf{BS2}, respectively. 

\subsection{Spin properties}
\subsubsection{Total Spin}
\protect\label{sec:spin}
\begin{table} [tbh!]
  \caption{Local Mulliken atomic spin densities, $N_\alpha - N_\beta$, on metal and oxygen atoms, 
           computed for the NiO and MnO with $5\times 5\times 5$ k-point grid. In the text, BS1 and BS2 refer to the broken-symmetry solutions obtained in cells 1 and 2, respectively.
\protect\label{tbl:MO_Mul_Sloc}}
\makebox[\textwidth]{
\begin{tabular}{c|cc|cc|cc|cc}
\hline
\hline
            & \multicolumn{4}{c}{\textbf{Cell 1}} & \multicolumn{4}{c}{\textbf{Cell 2}}  \\ 
\hline
            & \multicolumn{2}{c|}{NiO} & \multicolumn{2}{c}{MnO} & \multicolumn{2}{c|}{NiO} & \multicolumn{2}{c}{MnO}\\ 
\hline
            &  Ni      & O      &  Mn    & O        &  Ni      & O      &  Mn    & O     \\
UHF, HS     &  1.868   & 0.132  & 4.940  & 0.060    &  1.868   & 0.132  &  4.940 & 0.060 \\
GW, HS      &  1.794   & 0.206  & 4.907  & 0.093    &  1.794   & 0.206  &  4.907 & 0.093 \\
\hline                                            
UHF, BS     &  1.850   & 0.046  & 4.973  & 0.022    &  1.816   & 0.000  &  4.937 & 0.000 \\
GW, BS      &  1.793   & 0.070  & 4.934  & 0.033    &  1.710   & 0.000  &  4.870 & 0.000 \\
\hline
\hline
\end{tabular}
}
\end{table}

Table~\ref{tbl:MO_Mul_Sloc} shows local Mulliken spin densities on each of the atoms\cite{Mulliken1955}. 
The HS solutions are symmetric---local spin densities on both metal centers in the unit cell are the same. 
The spin-densities on the oxygen atoms are non-zero and are also symmetric. 
The BS solutions have the opposite signs of spin-density on different metal and oxygen centers. 

The fact that the oxygen spin density is nonzero demonstrates that oxygen is involved in magnetic interactions.  
This also indicates that care should be exercised when employing local approximations such as the ones in Ref. \cite{Kotani:SF-GW:2008}. In these local treatments,   
oxygen's contributions may not be accurately accounted for. 
Moreover, for HS and BS1 (broken-symmetry solution in cell \textbf{1}) solutions, we see an increase of spin-polarization on oxygen from UHF to GW, 
suggesting that oxygen's electrons are significantly correlated. 
Local spin-densities on metal centers for all the solutions are close to $2$ and $5$ for NiO and MnO, respectively. 
The Mulliken analysis is known to have deficiencies\cite{Baker:Mulliken:problems:1985},  
but it can be successfully used to understand trends of these solutions since the basis set employed by us is compact 
and free of linear dependencies.   

\begin{table} [tbh!]
  \caption{Total $\braket{S^2}$ (a.u.), computed for NiO, \textbf{cell 1}. 
\protect\label{tbl:NiO_90_S2tot}}
\makebox[\textwidth]{
\begin{tabular}{c|cccc}
\hline
\hline
            & $2\times 2\times 2$ & $3\times 3\times 3$ & $4\times 4\times 4$ & $5\times 5\times 5$ \\
UHF, HS     & 272.089             & 2970.295            & 16512.688           & 62751.341 \\
GW disc, HS & 276.448             & 2985.821            & 16550.047           & 62824.715 \\
ideal  HS   & 272                 & 2970                & 16512               & 62750 \\
\hline
            & $2\times 2\times 2$ & $3\times 3\times 3$ & $4\times 4\times 4$ & $5\times 5\times 5$ \\
UHF, BS     & 8.443               & 54.262              & 128.612             & 251.193 \\
GW disc, BS & 14.715              & 69.773              & 165.938             & 324.494 \\
ideal  BS   & 16                  & 54                  & 128                 & 250 \\
\hline
\hline
\end{tabular}
}
\end{table}
\begin{table} [tbh!]
  \caption{Total $\braket{S^2}$ (a.u.), computed for NiO, \textbf{cell 2}. 
\protect\label{tbl:NiO_60_S2tot}}
\makebox[\textwidth]{
\begin{tabular}{c|cccc}
\hline
\hline
            & $2\times 2\times 2$ & $3\times 3\times 3$ & $4\times 4\times 4$ & $5\times 5\times 5$ \\
UHF, HS     & 272.097             & 2970.299            & 16512.689           & 62751.341 \\
GW disc, HS & 276.469             & 2985.796            & 16549.919           & 62824.194 \\
ideal  HS   & 272                 & 2970                & 16512               & 62750 \\
\hline
            & $2\times 2\times 2$ & $3\times 3\times 3$ & $4\times 4\times 4$ & $5\times 5\times 5$ \\
UHF, BS     & 16.009              & 54.133              & 128.337             & 250.667  \\
GW disc, BS & 20.061              & 68.930              & 164.095             & 320.916  \\
ideal  BS   & 16                  & 54                  & 128                 & 250 \\
\hline
\hline
\end{tabular}
}
\end{table}
\begin{table} [tbh!]
  \caption{Total $\braket{S^2}$ (a.u.), computed for MnO, \textbf{cell 1}.
\protect\label{tbl:MnO_90_S2tot}}
\makebox[\textwidth]{
\begin{tabular}{c|cccc}
\hline
\hline
            & $2\times 2\times 2$ & $3\times 3\times 3$ & $4\times 4\times 4$ & $5\times 5\times 5$ \\
UHF, HS     & 1640.057            & 18360.198           & 102720.469          & 391250.917 \\
GW disc, HS & 1643.675            & 18373.045           & 102751.374          & 391311.545 \\            
ideal  HS   & 1640                & 18360               & 102720              & 391250 \\ 
\hline
            & $2\times 2\times 2$ & $3\times 3\times 3$ & $4\times 4\times 4$ & $5\times 5\times 5$ \\
UHF, BS     & 40.019              & 135.080             & 320.199             & 624.521 \\
GW disc, BS & 43.539              & 147.597             & 350.310             & 682.269 \\
ideal  BS   & 40                  & 135                 & 320                 & 625    \\
\hline
\hline
\end{tabular}
}
\end{table}
\begin{table} [tbh!]
  \caption{Total $\braket{S^2}$ (a.u.), computed for MnO, \textbf{cell 2}.
\protect\label{tbl:MnO_60_S2tot}}
\makebox[\textwidth]{
\begin{tabular}{c|cccc}
\hline
\hline
            & $2\times 2\times 2$ & $3\times 3\times 3$ & $4\times 4\times 4$ & $5\times 5\times 5$ \\
UHF, HS     & 1640.057            & 18360.197           & 102720.468          & 391250.917 \\
GW disc, HS & 1643.648            & 18372.999           & 102751.306          & 391311.545 \\
ideal  HS   & 1640                & 18360               & 102720              & 391250 \\
\hline
            & $2\times 2\times 2$ & $3\times 3\times 3$ & $4\times 4\times 4$ & $5\times 5\times 5$ \\
UHF, BS     & 39.945              & 134.888             & 319.751             & 624.521 \\
GW disc, BS & 43.302              & 147.067             & 349.133             & 682.269 \\
ideal  BS   & 40                  & 135                 & 320                 & 625    \\
\hline
\hline
\end{tabular}
}
\end{table}

Values of total $\braket{S^2}$ computed for all the solutions for different number of the k-points are shown in Tables~\ref{tbl:NiO_90_S2tot},\ref{tbl:NiO_60_S2tot},\ref{tbl:MnO_90_S2tot},\ref{tbl:MnO_60_S2tot}. 
We also show values of $\braket{S^2}$ for the model Ising configurations (determinants) 
in order to compare to the ideal $\braket{S^2}$ values. 
All the obtained HS UHF solutions are almost pure by spin, 
showing only minor deviations from the ideal $\braket{S^2}$ values.  
All broken-symmetry UHF solutions, with the exception of NiO BS1 UHF solution for the $2\times 2\times 2$ k-point grid, show minor deviations from the ideal values of the corresponding Ising configurations;  
the deviations are similar in magnitudes to the spin contamination of the HS solutions. 
For NiO, the BS1 UHF solution for the $2\times 2\times 2$ k-point grid (see Table~\ref{tbl:NiO_90_S2tot}) displays a large deviation from the ideal broken-symmetry $\braket{S^2}$ value, 
indicating that severe finite-size effects led to a solution with a different character. 
The absolute values of deviations increase with the size of the k-point grid. 

The disconnected component of the GW two-particle density matrix shows a residual spin contamination (which is present even for closed-shell systems\cite{Pokhilko:tpdm:2021}).
However, even despite this contamination, all the trends present in UHF solutions are also observed in GW solutions as well. 

\subsubsection{Values of $SS_u^{\mathbf{k}=0}$ and $SS_u$}

\begin{figure}[!h]
  \includegraphics[width=7cm]{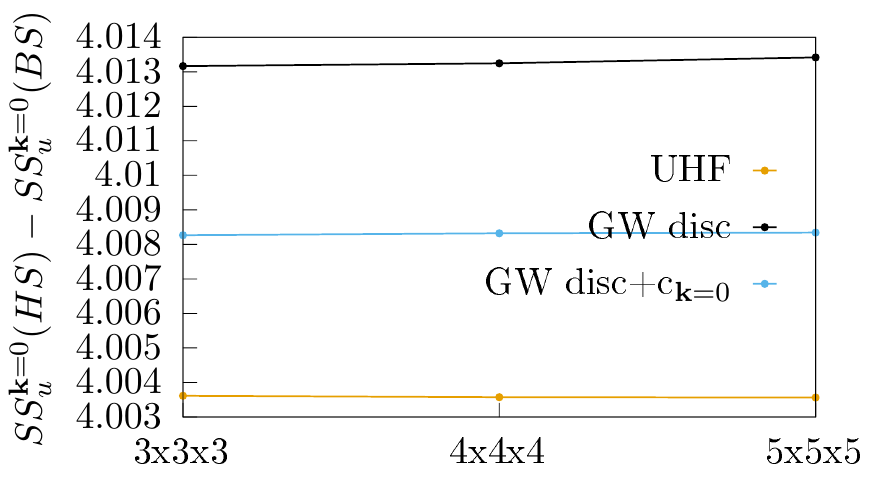}
  \includegraphics[width=7cm]{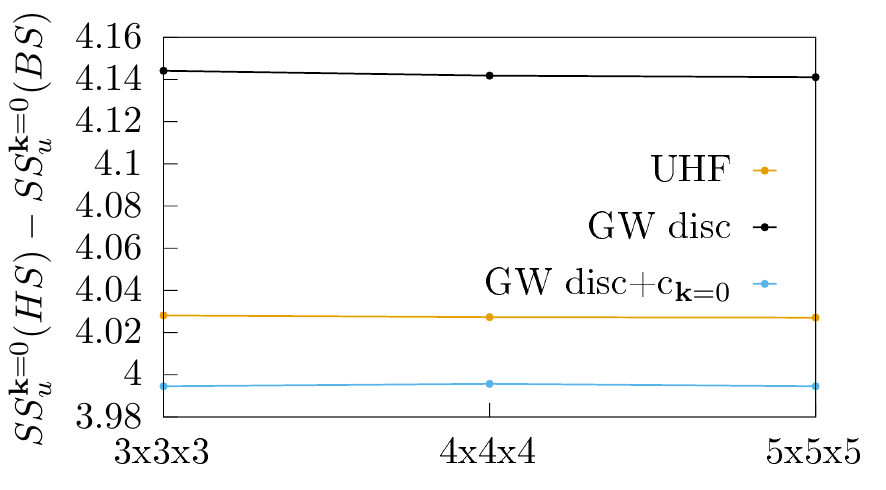}
\centering
\caption{Differences of $SS_u^{\mathbf{k}=0}$ between the HS and BS solutions,
         computed with UHF and GW, evaluated at the $\Gamma$ point for NiO \textbf{cell 1} (left) and \textbf{cell 2} (right). 
         GW includes either only the disconnected term abbreviated as \emph{disc} or 
         both the disconnected and connected terms (only at zero momentum transfer) 
         abbreviated as \emph{disc+c$_{\mathbf{k}=0}$}. 
         \protect\label{fig:S2u_diff_gamma}
}
\end{figure}

\begin{figure}[!h]
  \includegraphics[width=7cm]{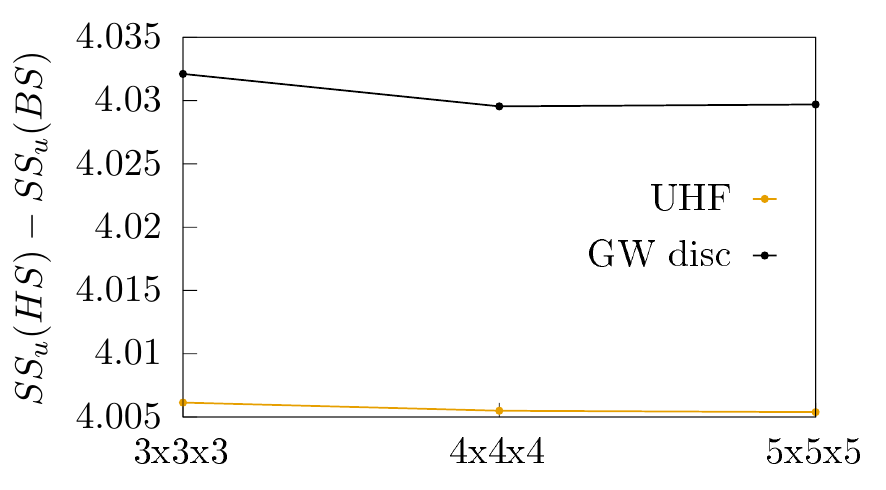}
  \includegraphics[width=7cm]{NiO_DSSu.pdf}
\centering
\caption{Differences of $SS_u$ between the HS and BS solutions  computed  with UHF and GW
         for NiO \textbf{cell 1} (left) and \textbf{cell 2} (right). 
         Only the disconnected contributions were used for GW. 
         \protect\label{fig:S2u_dif}
}
\end{figure}

\begin{figure}[!h]
  \includegraphics[width=7cm]{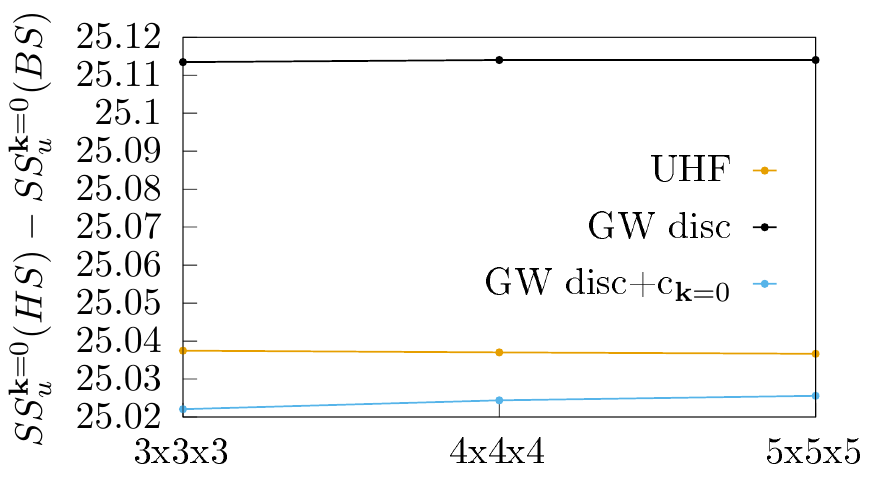}
  \includegraphics[width=7cm]{MnO_DS2gamma.pdf}
\centering
\caption{Differences of $SS_u^{\mathbf{k}=0}$ between the HS and BS solutions, 
         computed with UHF and GW, evaluated at the $\Gamma$  point for MnO \textbf{cell 1} (left) and \textbf{cell 2} (right). 
         GW includes either only the disconnected term abbreviated as \emph{disc} 
         or both the disconnected and connected terms (only at zero momentum transfer) 
         abbreviated as \emph{disc+c$_{\mathbf{k}=0}$}. 
         \protect\label{fig:MnO_S2u_diff_gamma}
}
\end{figure}

\begin{figure}[!h]
  \includegraphics[width=7cm]{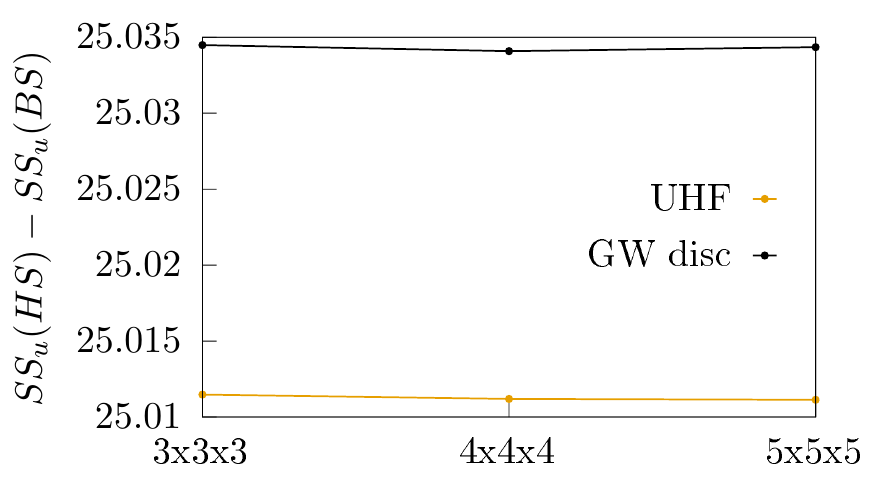}
  \includegraphics[width=7cm]{MnO_DSSu.pdf}
\centering
\caption{Differences of $SS_u$ between the HS and BS solutions, 
         computed for MnO \textbf{cell 1} (left) and \textbf{cell 2} (right) with UHF and GW. 
         Only disconnected contributions were used for GW. 
         \protect\label{fig:MnO_S2u_dif}
}
\end{figure}

It is hard to understand how the spin contamination changes with the size of the k-point grid since 
total $\braket{S^2}$ is not an extensive property. 
Consequently, we move to discussing extensive correlators $SS_u^{\mathbf{k}=0}$ and $SS_u$ described in Sec.~\ref{sec:SSuk} and Sec.~\ref{sec:SSu}, respectively.
Figures~\ref{fig:S2u_diff_gamma} and \ref{fig:MnO_S2u_diff_gamma}
show differences of $SS_u^{\mathbf{k}=0}$ between the HS and BS solutions for all cells. 
Similarly, Figures~\ref{fig:S2u_dif} and \ref{fig:MnO_S2u_dif} 
show differences of $SS_u$ between the HS and BS solutions for all cells. 
In both cases, we see that these differences change very little with the size of the k-grid, 
remaining close to the ideal values of $4$ and $25$ for NiO and MnO, respectively. 
A good agreement between $SS_u$ and $SS_u^{\mathbf{k}=0}$ suggests that 
the BS and HS solutions differ only locally 
(each cell in the real space representation contributes with the same high-spin or broken-spin characters) 
and no delocalization happens as in metals, where conducting electrons lead to RRKY interaction. 
This observation confirms the local structure of the effective magnetic Hamiltonian and 
the independence of the effective exchange constant $J$ from the momentum grid.  
To investigate the impact of the electronic cumulant, we tested different $SS_u^{\mathbf{k}=0}$ values  evaluated with partial inclusion of the cumulant 
with the zero momentum transfer as shown in Figures~\ref{fig:S2u_diff_gamma} and \ref{fig:MnO_S2u_diff_gamma}.
We observed that  the differences in $SS_u^{\mathbf{k}=0}$ became even closer to the ideal values of Ising configurations 
and even closer than the UHF results, despite displaying additional residual spin contamination in the absolute values of $SS_u^{\mathbf{k}=0}$.
This result illustrates that the cumulant not only preserves the structure of the solutions given 
by the disconnected contributions, but also makes the estimate closer to the idealized Ising configurations. 

\subsubsection{Values of $SS_{AB}^{\mathbf{k}=0}$ and $SS_{AB}$}
\begin{figure}[!h]
  \includegraphics[width=7cm]{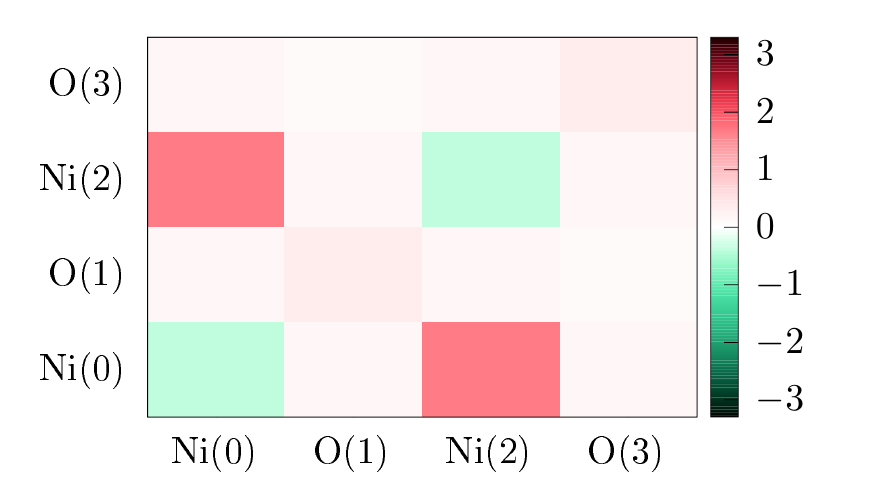}
  \includegraphics[width=7cm]{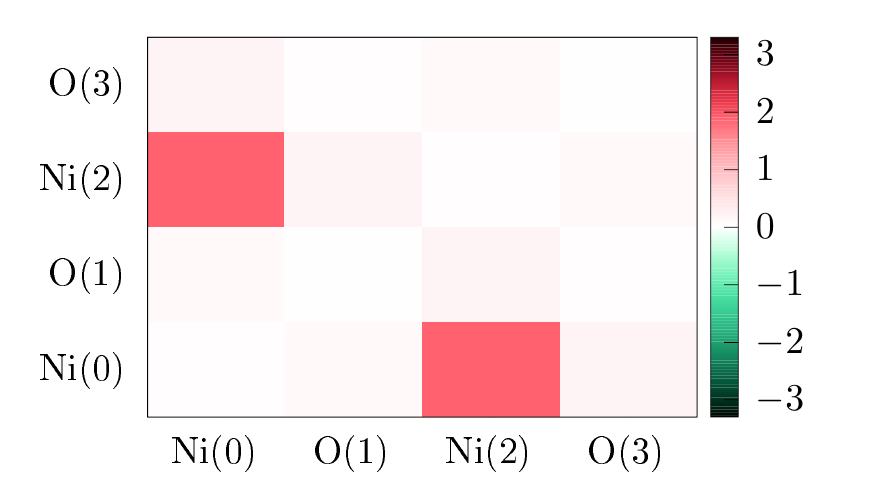} \\
  \includegraphics[width=7cm]{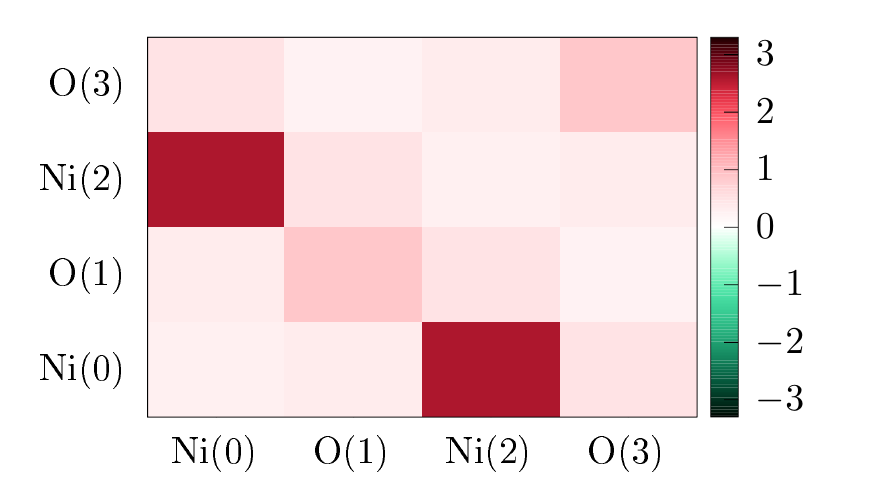}
  \includegraphics[width=7cm]{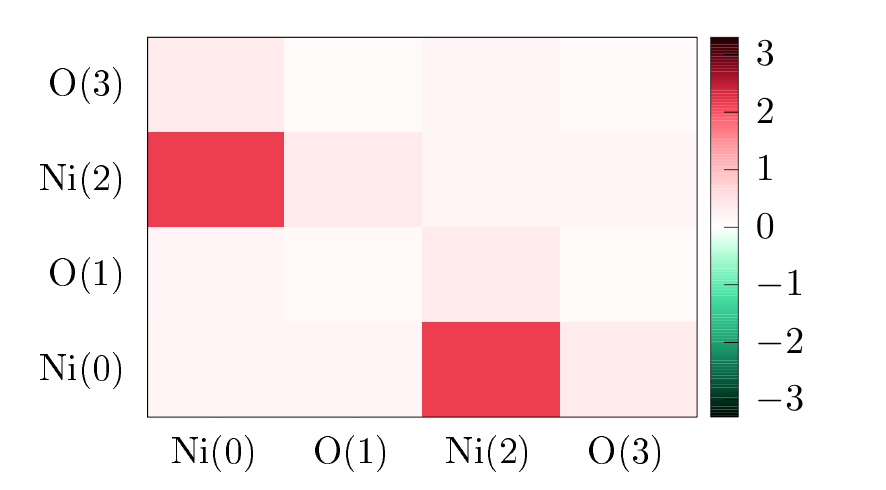} 
\centering
\caption{Differences in local spin--spin correlators between HS and BS2 solutions for NiO, \textbf{cell 2} 
         within $5\times 5 \times 5$ k-point grid. 
         The differences in the left column are evaluated only for $\mathbf{k}=0$ ($SS_{AB}^{\mathbf{k}=0}$); 
         the differences in the right column are evaluated for all $\mathbf{k}$ ($SS_{AB}$). 
         The plots at the top are evaluated with UHF; the plots at the bottom are evaluated with GW. 
         Only disconnected contributions are used for GW correlators. 
         \protect\label{fig:NiO_60_Sloc_dif}
}
\end{figure}
\begin{figure}[!h]
  \includegraphics[width=7cm]{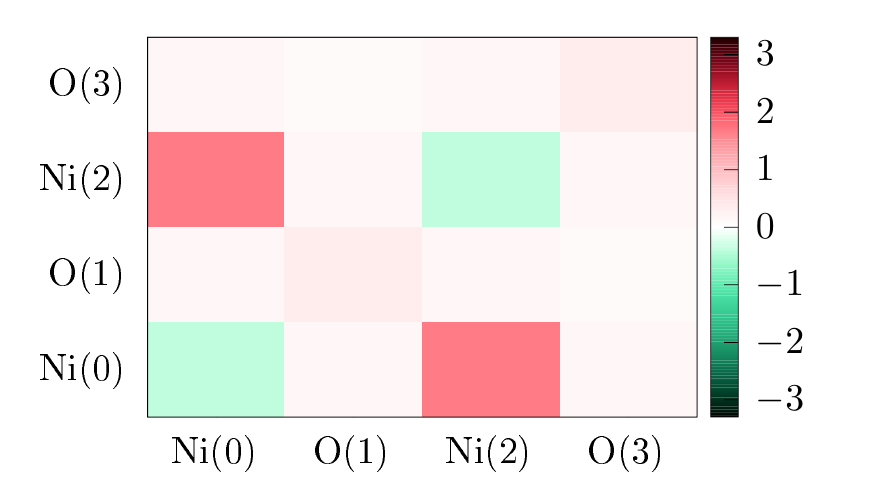}
  \includegraphics[width=7cm]{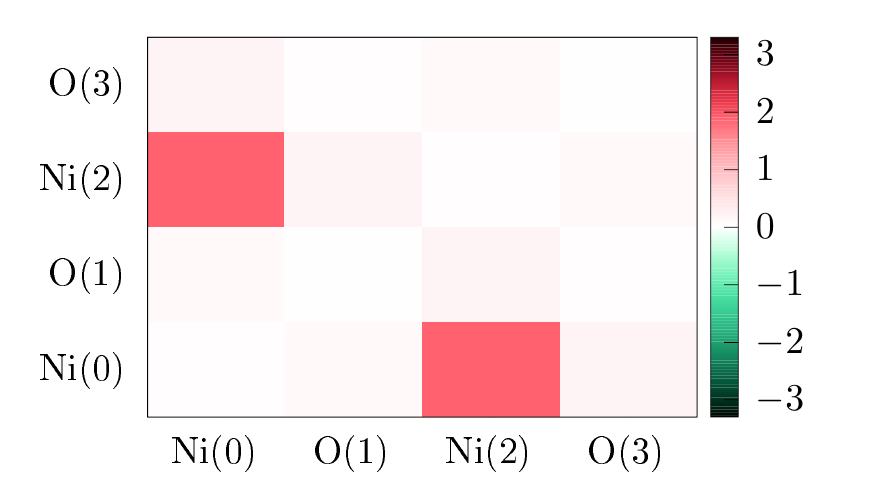} \\
  \includegraphics[width=7cm]{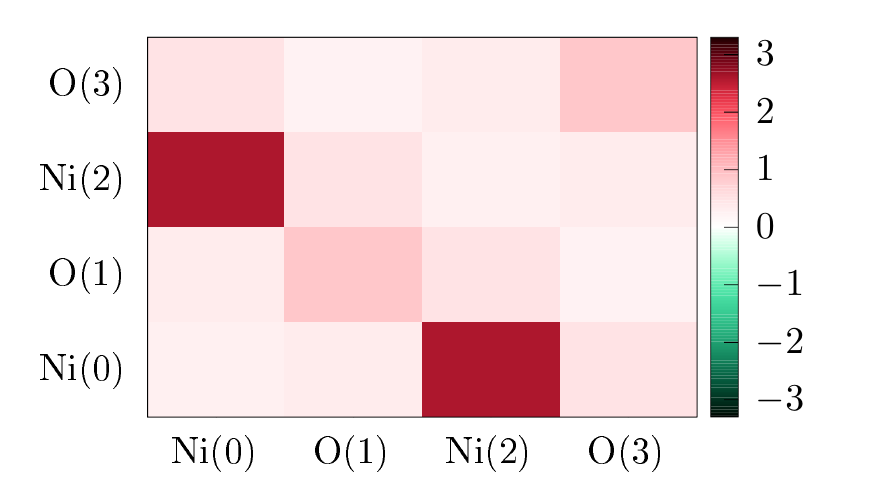}
  \includegraphics[width=7cm]{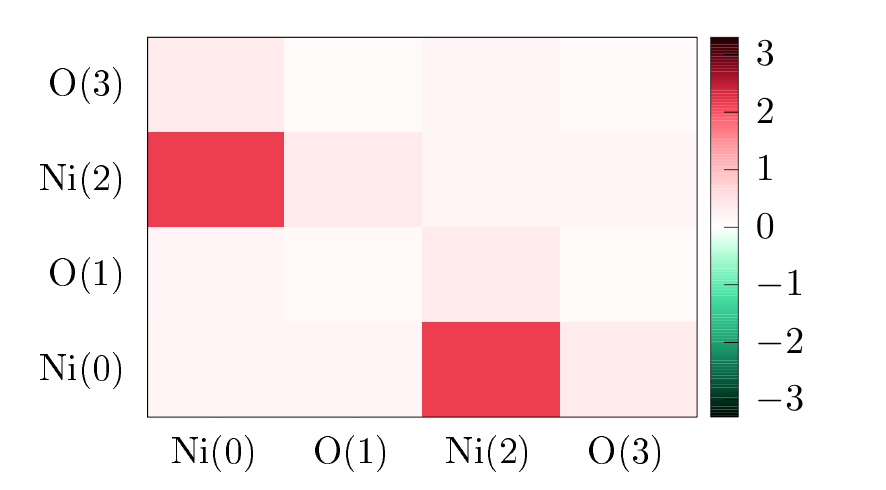} 
\centering
\caption{Differences in local spin--spin correlators between HS and BS1 solutions for NiO, \textbf{cell 1} 
         within $5\times 5 \times 5$ k-point grid. 
         The differences in the left column are evaluated only for $\mathbf{k}=0$ ($SS_{AB}^{\mathbf{k}=0}$); 
         the differences in the right column are evaluated for all $\mathbf{k}$ ($SS_{AB}$). 
         The plots at the top are evaluated with UHF; the plots at the bottom are evaluated with GW. 
         Only disconnected contributions are used for GW correlators. 
         \protect\label{fig:NiO_90_Sloc_dif}
}
\end{figure}
\begin{figure}[!h]
  \includegraphics[width=7cm]{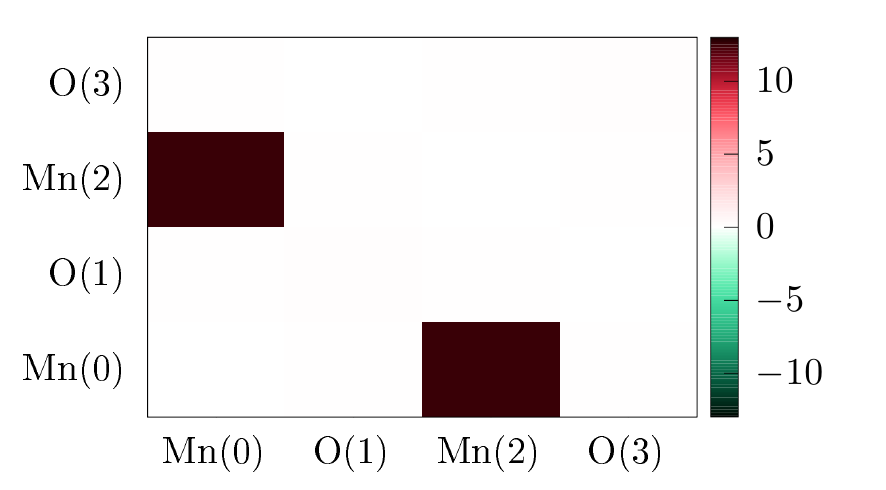}
  \includegraphics[width=7cm]{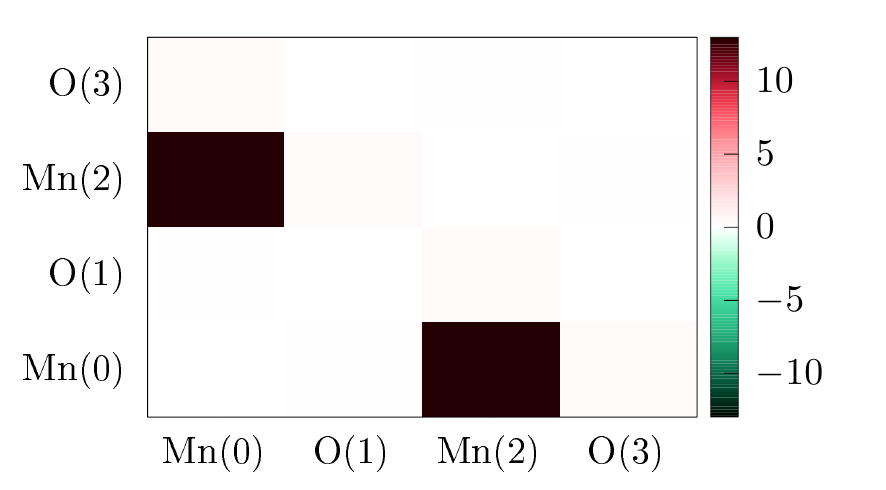} \\
  \includegraphics[width=7cm]{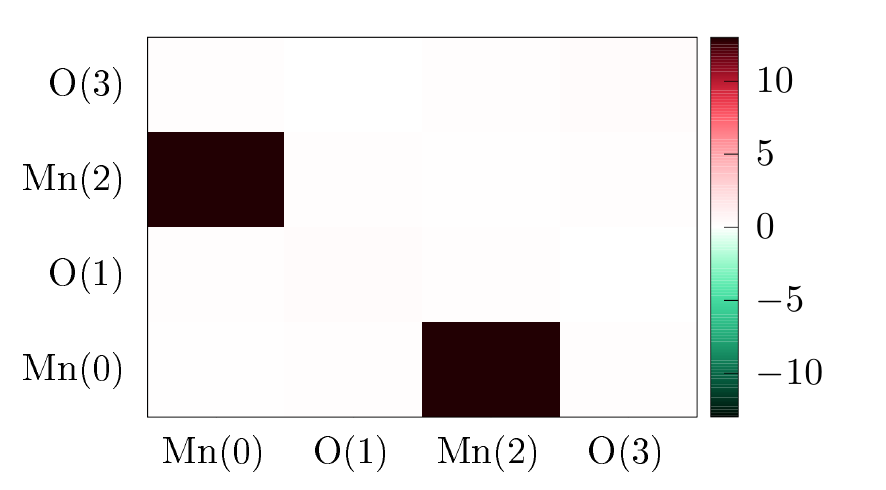}
  \includegraphics[width=7cm]{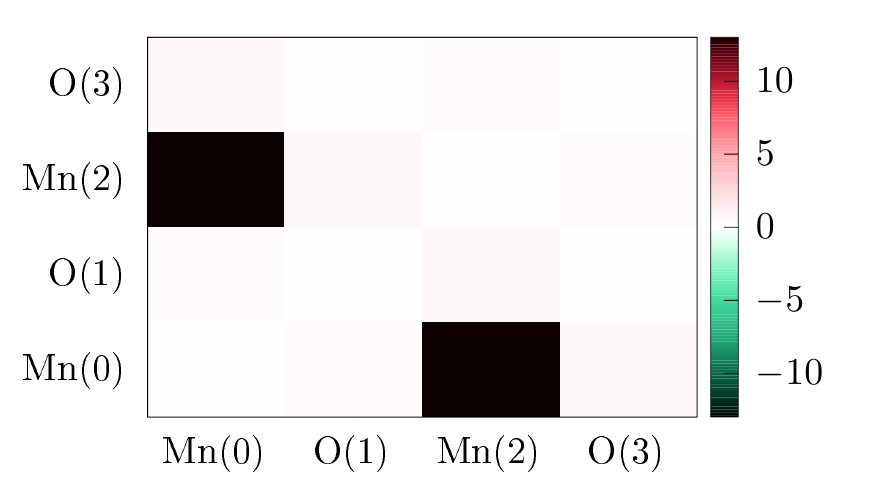} 
\centering
\caption{Differences in local spin--spin correlators between HS and BS2 solutions for MnO, \textbf{cell 2} 
         within $5\times 5 \times 5$ k-point grid. 
         The differences in the left column are evaluated only for $\mathbf{k}=0$ ($SS_{AB}^{\mathbf{k}=0}$); 
         the differences in the right column are evaluated for all $\mathbf{k}$ ($SS_{AB}$). 
         The plots at the top are evaluated with UHF; the plots at the bottom are evaluated with GW. 
         Only disconnected contributions are used for GW correlators. 
         \protect\label{fig:MnO_60_Sloc_dif}
}
\end{figure}
\begin{figure}[!h]
  \includegraphics[width=7cm]{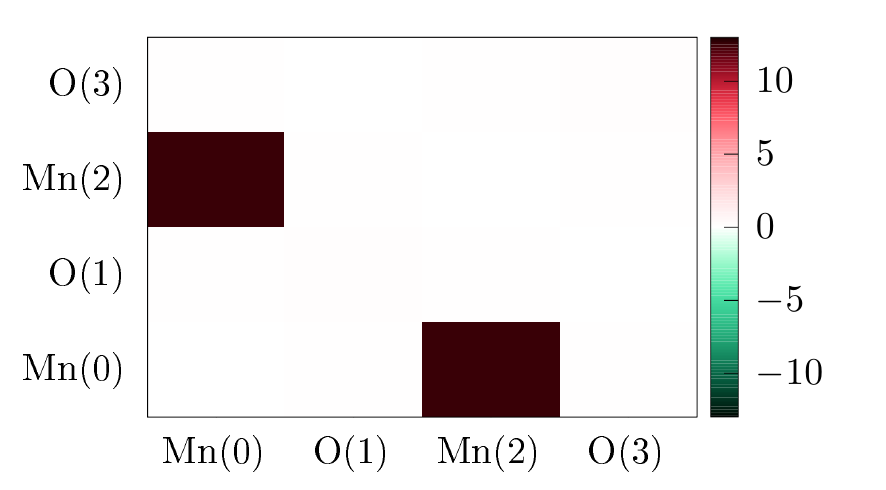}
  \includegraphics[width=7cm]{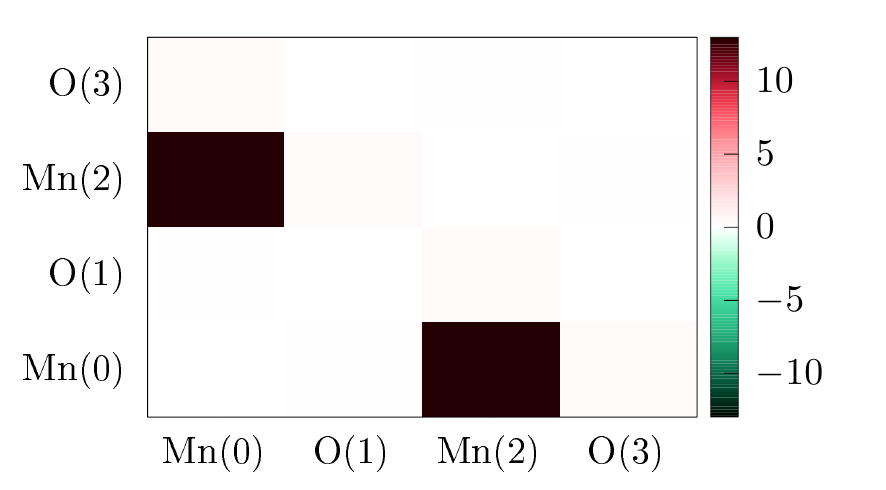} \\
  \includegraphics[width=7cm]{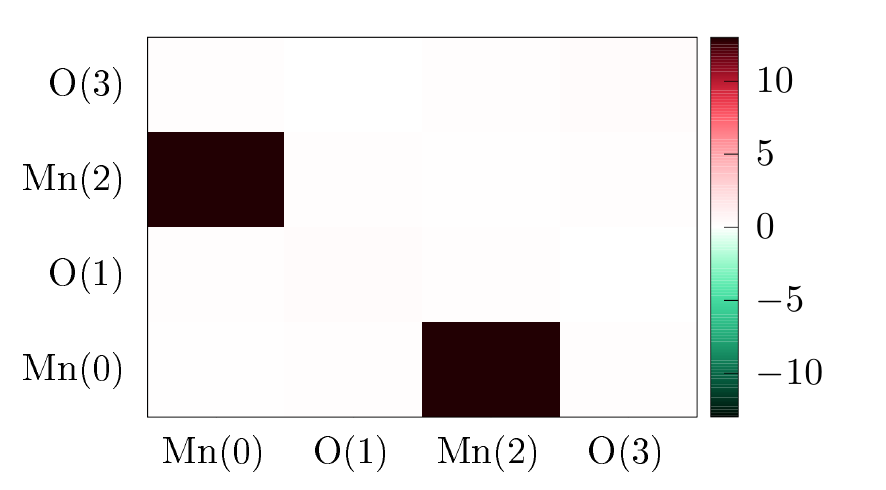}
  \includegraphics[width=7cm]{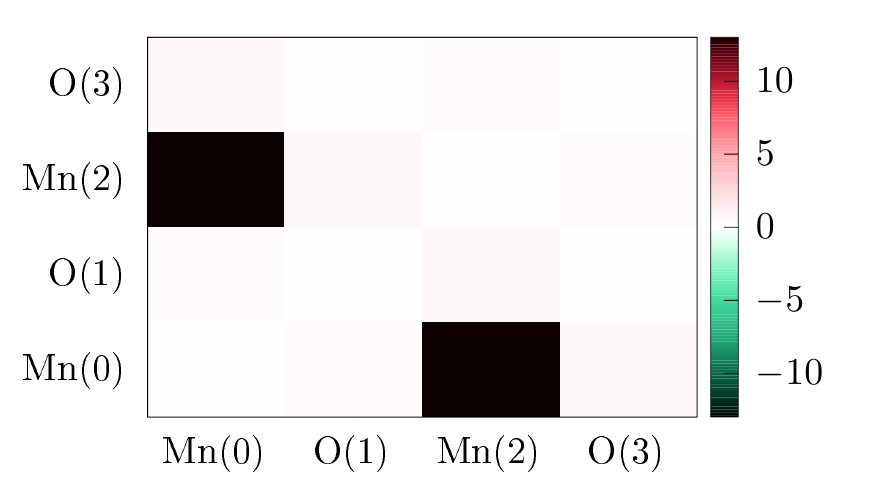} 
\centering
\caption{Differences in local spin--spin correlators between HS and BS1 solutions for MnO, \textbf{cell 1} 
         within $5\times 5 \times 5$ k-point grid. 
         The differences in the left column are evaluated only for $\mathbf{k}=0$ ($SS_{AB}^{\mathbf{k}=0}$); 
         the differences in the right column are evaluated for all $\mathbf{k}$ ($SS_{AB}$). 
         The plots at the top are evaluated with UHF; the plots at the bottom are evaluated with GW. 
         Only disconnected contributions are used for GW correlators. 
         \protect\label{fig:MnO_90_Sloc_dif}
}
\end{figure}
Figures~\ref{fig:NiO_60_Sloc_dif},\ref{fig:NiO_90_Sloc_dif},\ref{fig:MnO_60_Sloc_dif},\ref{fig:MnO_90_Sloc_dif} 
show differences between local spin--spin correlators ($\Delta SS_{AB}^{\mathbf{k}=0}$ and $\Delta SS_{AB}$) 
evaluated for the HS and BS solutions for NiO and MnO. 
The numerical values of each of the correlators are shown in Section~1 in SI. 
Although the HS and BS solutions differ mostly by the local spin orientation on metal atoms, 
the local spin correlators clearly show that there are additional features. 
First, the $\Delta SS_{AB}^{\mathbf{k}=0}$ on the same metal centers are non-zero, 
meaning that the local spins of the HS and BS solutions are slightly different. 
Additionally, the  $\Delta SS_{AB}$ on oxygen atoms are non-zero. 
These features indicate that the degree of spin polarization of oxygen varies for different solutions, 
which is consistent with the one-electron Mulliken analysis of spin densities in Table~\ref{tbl:MO_Mul_Sloc}. 
Due to limitations of the Mulliken analysis,  caution should exercised when  interpreting 
 numerical values of the local correlators. 
For example, $\Delta SS_{M_1 M_2}^{\mathbf{k}=0}$ between two different magnetic centers $M_1$ and $M_2$ of an 
ideal system with local spins $S=1$ equals to 2. 
The UHF $\Delta SS_{Ni_1 Ni_2}^{\mathbf{k}=0}$ for NiO, \textbf{cell 2}, is 3.280. 
Such a large deviation \emph{shall not} be interpreted as a signature of spin contamination 
because it heavily relies on the spin locality and on a particular choice of the localization 
(we use Mulliken partitioning). 
This reasoning is confirmed by noticing that UHF $\Delta SS_u^{\mathbf{k}=0}$ evaluated for the entire unit cell is 4.027, while the value from the idealized Ising configurations is 4.

\subsection{Exchange constants}
\begin{figure}[!h]
  \includegraphics[width=7cm]{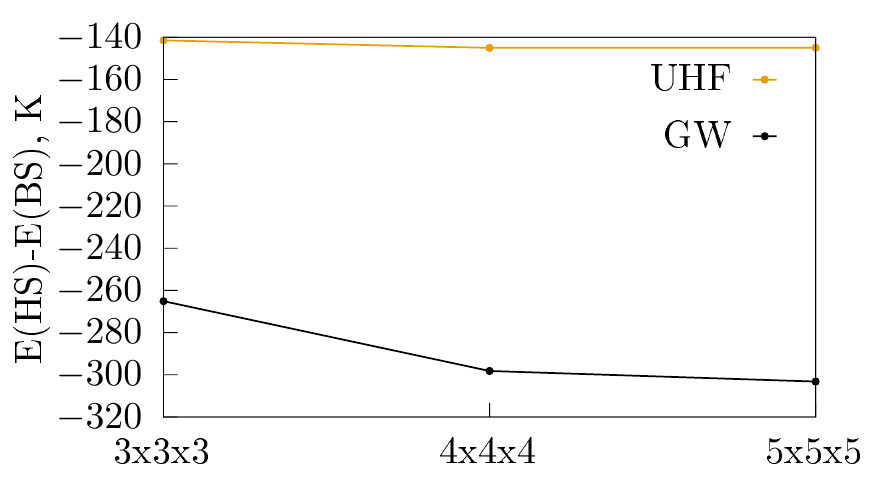}
  \includegraphics[width=7cm]{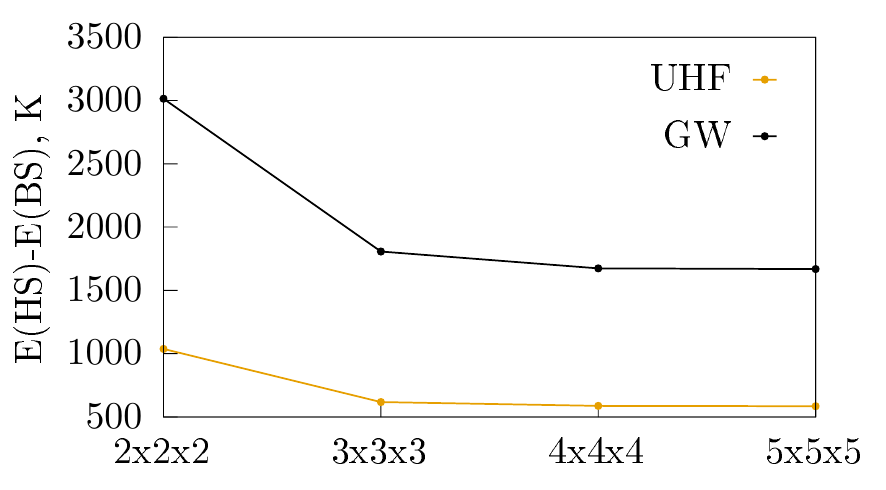}
\centering
\caption{Energy differences between the solutions, computed for NiO \textbf{cell 1} (left) 
         and \textbf{cell 2} (right) cells.
         \protect\label{fig:NiO_DE}
}
\end{figure}
\begin{figure}[!h]
  \includegraphics[width=7cm]{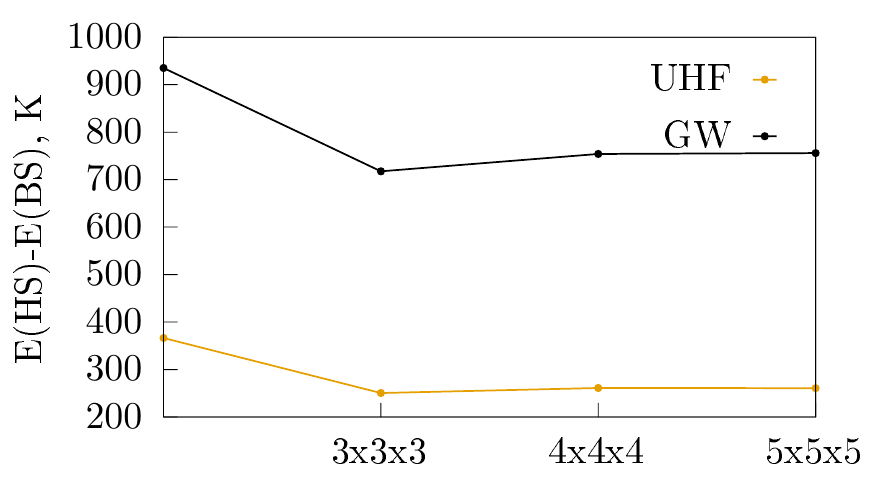}
  \includegraphics[width=7cm]{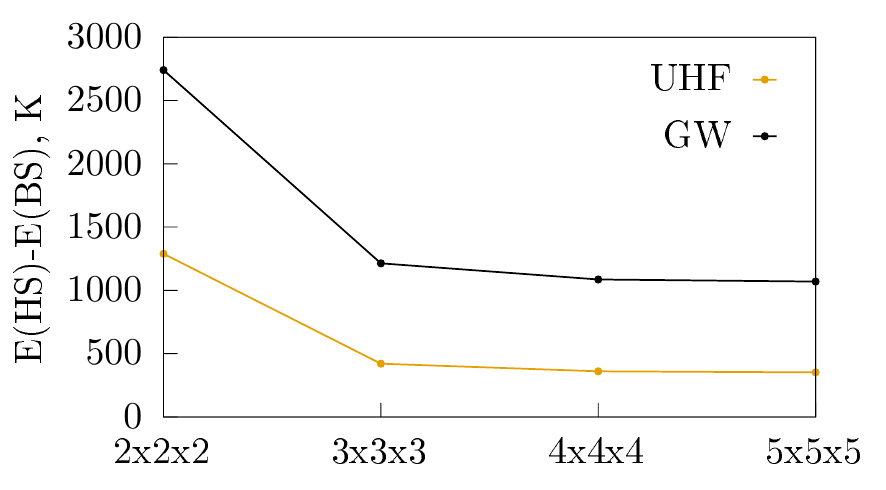}
\centering
\caption{Energy differences between the solutions, computed for MnO \textbf{cell 1} (left) and \textbf{cell 2} (right).
         \protect\label{fig:MnO_DE}
}
\end{figure}
In SI, Tables~17,18,20,21 show 
total energies per unit cell computed at the UHF and GW levels for all the solutions. 
The convergence of total energies with the number of k-points is relatively slow. 
For example, the total energies of the HS NiO solutions computed for \textbf{cell 1} and \textbf{cell 2} are expected to be the same at TDL; however, 
with $5\times 5 \times 5$ k-point grid they are still different by more than $0.03$ Hartree.
Unlike total energies, the energy differences converge very rapidly with the size of the k-point grid, as
shown in Figures~\ref{fig:NiO_DE} and \ref{fig:MnO_DE}. 
This is expected since the orbital occupancies of the HS and BS solutions are similar,  
resulting in cancellations of the finite-size errors. 
The only deviation from this trend is given by the BS1 $2\times 2\times 2$ solution for NiO, 
showing a big energy deviation, 
which is predicted by the $\braket{S^2}$ diagnostic in Section~\ref{sec:spin}. 
UHF estimates converge slightly faster the GW ones, 
but already at the $5\times 5 \times 5$ the energy differences are fully converged even at the GW level. 
\begin{figure}[!h]
  \includegraphics[width=7cm]{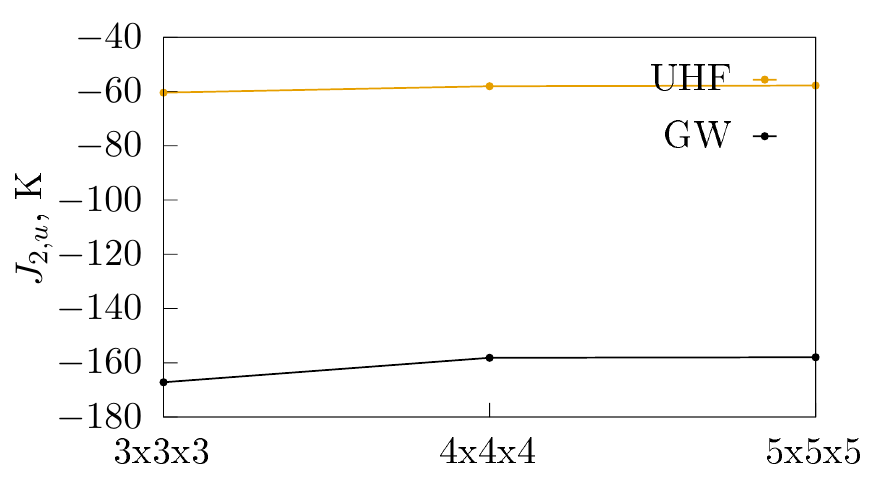}
  \includegraphics[width=7cm]{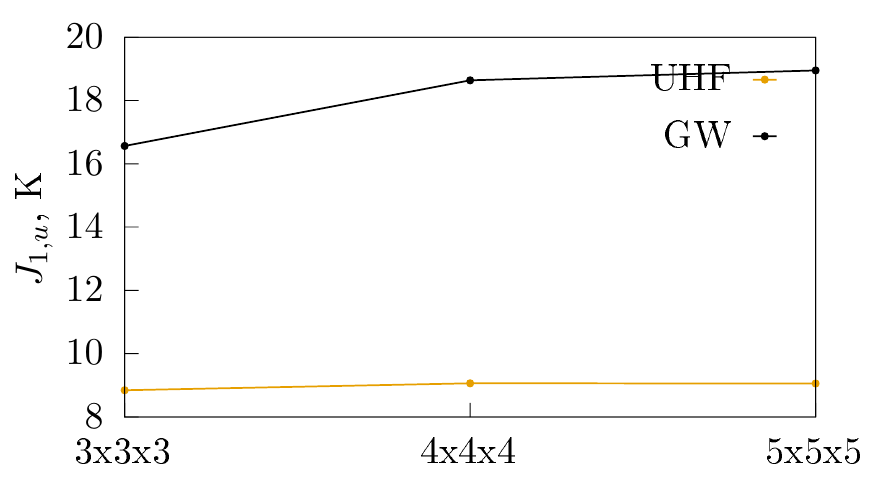}
\centering
\caption{Convergence of effective exchange couplings for NiO. 
         \protect\label{fig:NiO_J}
}
\end{figure}
\begin{figure}[!h]
  \includegraphics[width=7cm]{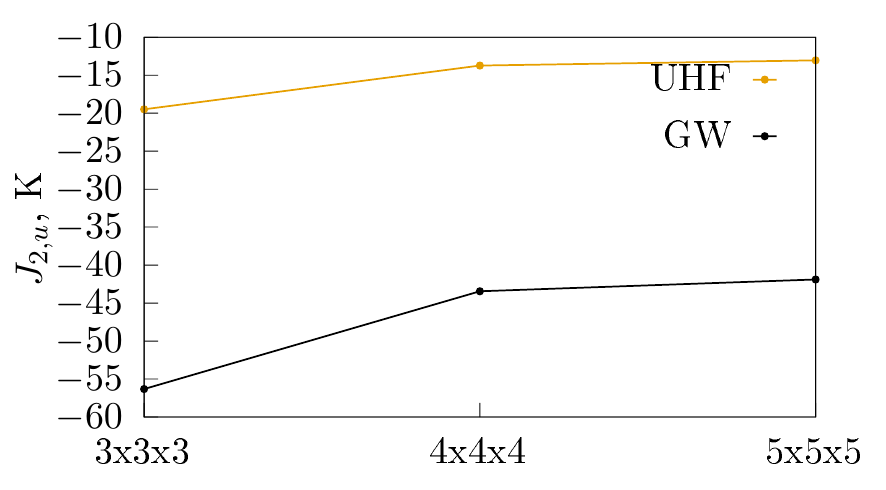}
  \includegraphics[width=7cm]{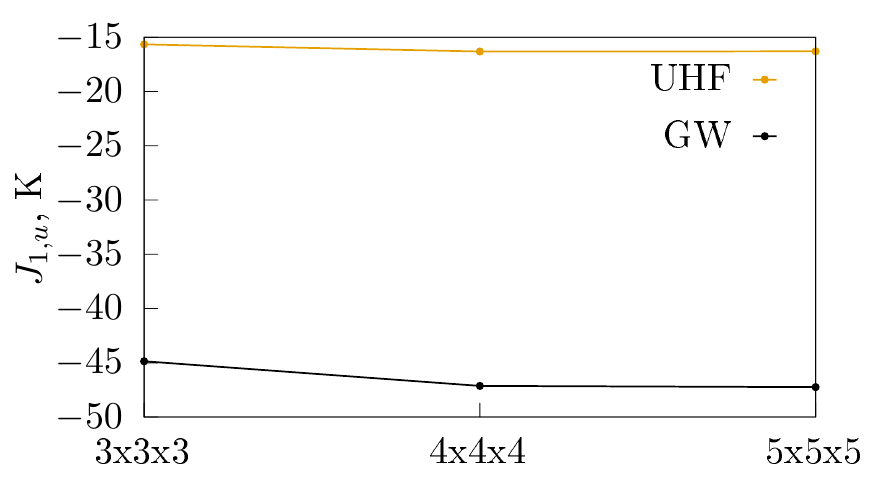}
\centering
\caption{Convergence of effective exchange couplings for MnO. 
         \protect\label{fig:MnO_J}
}
\end{figure}

The lack of strong dependence of $\Delta SS_u$ and $\Delta SS_u^{\mathbf{k}=0}$ on the k-grid size,
similarity in their numerical values, and rapid convergence of the energy differences 
suggest a local structure of the magnetic Hamiltonian, 
confirming the expression given in Eq.~\ref{eq:Ham_defs}. 
Our extraction of the effective exchange couplings is slightly different from  Ref.\cite{Majumdar:NiO:MnO:DFT:J:2011} 
since we use two cells
\begin{gather}
J_{1,u} = -\frac{E(HS1)-E(BS1)}{2\cdot 8}, \\
J_{2,u} = -\frac{E(HS2)-E(BS2)}{2\cdot 6} - J_{1,u}
\protect\label{eq:J_extr}
\end{gather}
where $E()$ are the total energies of the corresponding solutions, 
$J_{n,u}$ are the effective exchange couplings per formula unit for unit spins.  
The division by $2$ is highlighted because the unit cells that we use have two formula units. 
Since spin contamination is very small and its definition is somewhat ambigous, 
we do not correct the extracted $J$ couplings for it. 
A rapid convergence in energy differences leads to the rapid convergence of the extracted $J$ couplings 
with respect to the k-point grid, as shown in Figures~\ref{fig:NiO_J} and \ref{fig:MnO_J} and 
Tables~19 and 22 in SI.  

\begin{table} [tbh!]
  \caption{Comparison between effective exchange couplings $J$, meV, evaluated in this work and previous publications. 
\protect\label{tbl:J_comparison}}
\makebox[\textwidth]{
\begin{tabular}{c|cc|c|cc}
\hline
\hline
            & \multicolumn{2}{c|}{NiO} & & \multicolumn{2}{c}{MnO}  \\ 
            & $J_{1,u}$  & $J_{2,u}$           & & $J_{1,u}$ & $J_{2,u}$            \\
\hline
            \multicolumn{6}{c}{This work}  \\ 
\hline
UHF         & 0.78   & -4.98           & UHF & -1.40 & -1.12 \\
GW          & 1.63   & -13.61          & GW  & -4.07 & -3.61 \\
\hline
            \multicolumn{6}{c}{Wave-function methods}  \\ 
\hline
UHF$^a$         & 0.8    & -4.6            & UHF$^d$    & -1.45   & -2.3 \\
CASSCF$^a$      & 0.5    & -5.0            & CASSCF$^e$ & -2.75   & -- \\
CASPT2$^a$      & 1.2    & -16.7           & CASPT2$^e$ & -8.19   & -- \\
DDCI2$^a$      & 1.2    & -12.6           &     &    & \\
DDCI3$^a$       & 1.8    & -16.3           &     &    & \\
\hline
            \multicolumn{6}{c}{DFT}  \\ 
\hline
LDA$^a$        & 11.9 & -71.3              &     &    &\\
B3LYP$^a$      & 2.4  & -26.7              & B3LYP$^b$    & -5.3   & -11.0 \\
PBE$^c$        & 1.2  & -44.5              & PBE$^c$      & -9.5   & -14.9 \\
PSIC$^c$       & 3.3  & -24.7              & PSIC$^c$     & -5.0   & -7.6  \\
HSE$^c$        & 2.3  & -21.0              & HSE$^c$      & -7.0   & -7.8  \\
ASIC$^c$       & 5.2  & -45.0              & ASIC$^c$     & -8.0   & -11.3 \\
PBE0$^d$       & -6.2   & -7.4             &       &   &\\
\hline
            \multicolumn{6}{c}{Green's function methods}  \\ 
\hline
qpGW+approx$^f$ & -0.8 & -14.7  & qpGW+approx & -2.8 & -4.7 \\
\end{tabular}
}

$^a$ Ref. \cite{Martin:NiO:exchange:2002} 

$^b$ Ref. \cite{Feng:MnO:CoO:b3lyp:2004}

$^c$ Ref. \cite{Majumdar:NiO:MnO:DFT:J:2011} 

$^d$ Ref. \cite{Kresse:MnO:PBE:2005}

$^e$ Ref. \cite{deGraaf:MRPT:exchange:solids:SMM:2001}

$^f$ qpGW with additional approximations from the Ref. \cite{Kotani:SF-GW:2008} 
\end{table}

Table~\ref{tbl:J_comparison} can be used to compare  the effective exchange couplings evaluated in this work at the UHF and GW level against the previous calculations.
Our UHF results for NiO are within an excellent agreement with 
the UHF estimates from  Ref.~\cite{Martin:NiO:exchange:2002}. 
Small discrepancies present are likely due to different basis sets and different treatments of finite-size effects. 
Our GW $J_1$ estimate is very close to an accurate DDCI3 estimate; 
the GW $J_2$ estimate is between the DDCI2 and DDCI3 estimates.  
This quality of the GW estimates is consistent with the performance of the GW for another oxo-bridged 
superexchange system---the Fe$_2$OCl$_6^{2-}$ complex, for which the GW estimate is between CASSCF and MRCI\cite{Pokhilko:local_correlators:2021,Morokuma:DMRG:biquad_exc:2014}. 
Our self-consistent GW estimate is somewhat similar to 
the quasiparticle GW estimate from Ref.\cite{Kotani:SF-GW:2008} for the $J_2$ coupling, 
but is qualitatively different for the $J_1$ coupling. 
While it is hard to pinpoint the source of this disagreement, 
it is likely due to several additional approximations used in the Ref.\cite{Kotani:SF-GW:2008}. 
 
For MnO, the $J_1$ UHF estimates from our periodic calculations agree well with Ref. \cite{deGraaf:MRPT:exchange:solids:SMM:2001}, 
but the $J_2$ estimates disagree. 
This disagreement is likely due to the finite-size effects present in molecular clusters employed in Ref.~\cite{deGraaf:MRPT:exchange:solids:SMM:2001}.
Since each manganese atom contains 5 unpaired electrons, 
the size of the molecular cluster that can be handled by multireference methods is severely limited by the size of the active space. 
It is likely that the size of the molecular cluster studied in  Ref.~\cite{deGraaf:MRPT:exchange:solids:SMM:2001}
was not sufficient for the NNN couplings $J_2$ to reach convergence to TDL. 
By the same reason the CASSCF and CASPT2 $J_2$ estimates are not reported. 
Moreover, in our periodic calculations we see that finite-size effects in MnO are more significant  than in NiO. This would mean that when finite clusters are used to evaluate $J$,  even larger clusters should be required for accurate estimates in MnO.
The GW $J_1$ estimate is between CASSCF and CASPT2 estimates. 
Our GW yields antiferromagnetic $J_1$ and $J_2$ constants, which qualitatively agree with qpGW from Ref.\cite{Kotani:SF-GW:2008}. 
However, unlike in Ref.\cite{Kotani:SF-GW:2008}, our estimates predict that the NN interactions in MnO are stronger than the NNN ones. 

The DFT estimates for NiO and MnO vary widely. 
Perhaps, the most realistic DFT estimates are provided by 
self-interaction-corrected functionals, such as HSE and PSIC from the Ref. \cite{Majumdar:NiO:MnO:DFT:J:2011}. 
However, a lack of systematic improvement makes it hard to assess the quality of these DFT calculations, 
especially when a direct comparison with experimental observables is not available.

\section{Conclusions}
We extended the evaluation of two-particle density matrices withing 
the one-particle Green's function formalism to periodic systems. 
We applied the evaluated two-particle density matrices to analyze electronic structure of nickel and manganese oxides.  
This extension allowed us to evaluate $\braket{S^2}$ values and investigate effects of spin contamination of 
the UHF and GW solutions for NiO and MnO. 
When total $\braket{S^2}$ is interpreted as an $\braket{S^2}$ of a supersystem, 
it can serve as a valuable diagnostic tool. 
In particular, $\braket{S^2}$ values quantified deviations from the ideal high-spin (ferromagnetic) 
and broken-symmetry (antiferromagnetic) configurations and demonstrated that for certain sizes of the k-point grid extracting the magnetic Hamiltonian was unreliable. To understand the influence of finite-size effects, 
we proposed two version of spin--spin correlation functions, $SS_u$ and $SS_u^{\mathbf{k}=0}$, which can be loosely 
interpreted as an ``$\braket{S^2}$ of a unit cell''. Both versions agree well with each other for both NiO and MnO demonstrating that the magnetic structure of these compounds is largely momentum independent. 
Although GW introduces an artificial spin contamination to the solutions, these artificial contributions cancel out 
when differences are considered. When the electronic cumulant is included, the estimates of $SS_u^{\mathbf{k}=0}$ become closer 
to the value of ideal Ising configurations making the extraction of effective magnetic Hamiltonians from 
these solutions unambiguous. 
Considering appropriate supercells, we extracted the effective magnetic couplings $J_1$ and $J_2$ between the NN and NNN centers in NiO and MnO. 
The resulting GW estimates agree very well with high-quality wave-function calculations. 
The values of the effective exchange couplings obtained at the GW level increase significantly when compared to UHF values.
This is likely due to an effective screening, captured by GW, which stabilizes charge transfer. 
In the upcoming work, we will focus on the mechanism of the magnetic interactions in MnO and NiO
 and provide a direct comparison with finite-temperature experiments.

\section*{Acknowledgments}
P.P. and D.Z. acknowledge support from the NSF grant CHE-1453894.

\section*{Supplementary Material}
Values of local spin correlators, total energies, and effective exchange constants. 

\section*{Data Availability}
The data that supports the findings of this study are available within the article [and its supplementary material].

\renewcommand{\baselinestretch}{1.5}


\begin{thebibliography}{100}
\bibitem{note:BS-AFM}
The term ``antiferromagnetic'' can be confusing, since it can also refer to the
  open-shell singlet configuration instead of the broken-spin configuration.
  The term ``phase'' can also cause an additional confusion for readers within
  molecular electronic structure. To minimize possible confusion and
  misinterpretation of the text, we explicitely name solutions by their
  broken-spin nature.
\end{thebibliography}
\end{document}